\PassOptionsToPackage{table}{xcolor}
\documentclass[acmsmall,screen]{acmart}
\AtBeginDocument{%
  }

\setcopyright{acmlicensed}
\copyrightyear{2018}
\acmYear{2018}
\acmDOI{XXXXXXX.XXXXXXX}

\acmJournal{JACM}
\acmVolume{37}
\acmNumber{4}
\acmArticle{111}
\acmMonth{8}

\usepackage{booktabs,multirow}
\usepackage{subcaption}
\usepackage{tcolorbox}       % For boxed prompts in appendix
\tcbuselibrary{breakable,skins}    % Allow boxes to break across pages
\usepackage{xspace}
\usepackage{enumitem}
\usepackage{algorithm}
\usepackage{algpseudocode}

\newtcolorbox{summarybox}[1]{
    colback=gray!5!white,
    colframe=gray!75!black,
    title=#1,
    fonttitle=\bfseries\small,
    enhanced,
    left=1mm,
    right=1mm,
    top=1mm,
    bottom=1mm
}

\newcommand{\rqzero}{Where are Bug-Inducing Commits Located?}
\newcommand{\rqone}{How Effective is TKG-Guided Agentic Search for BIC Identification?}
\newcommand{\rqtwo}{What is the Contribution of Each AgenticSZZ Component?}
\newcommand{\rqthree}{How Does LLM Choice Affect \appname's Effectiveness?}

\newcommand{\appname}{AgenticSZZ\xspace}

\begin{document}

\title{AgenticSZZ: Temporal Knowledge Graph-Guided Agentic Bug-Inducing Commit Identification}

\author{Yu Shi}
\affiliation{%
    \institution{Queen's University}
    \city{Kingston}
    \country{Canada}}
\email{y.shi@queensu.ca}
\orcid{0009-0005-6083-0932}

\author{Hao Li}
\affiliation{%
    \institution{Queen's University}
    \city{Kingston}
    \country{Canada}}
\email{hao.li@queensu.ca}
\orcid{0000-0003-4468-5972}

\author{Bram Adams}
\affiliation{%
    \institution{Queen's University}
    \city{Kingston}
    \country{Canada}}
\email{bram.adams@queensu.ca}
\orcid{0000-0001-7213-4006}

\author{Ahmed E. Hassan}
\affiliation{%
    \institution{Queen's University}
    \city{Kingston}
    \country{Canada}}
\email{hassan@queensu.ca}
\orcid{0000-0001-7749-5513}

\renewcommand{\shortauthors}{Yu Shi et al.}

\begin{abstract}
    Identifying Bug-Inducing Commits (BICs) is fundamental for understanding software defects and enabling downstream tasks such as defect prediction and automated program repair. Yet existing SZZ-based approaches are limited by their reliance on \texttt{git blame}, which restricts the search space to commits that directly modified the fixed lines. Our preliminary study on 2,102 validated bug-fixing commits reveals that this limitation is significant: over 40\% of cases cannot be solved by blame alone, as 28\% of BICs require traversing commit history beyond blame results and 14\% are blameless.

    We present \appname, the first approach to apply Temporal Knowledge Graphs (TKGs) to software evolution analysis. \appname reframes BIC identification from a ranking problem over blame commits into a graph search problem, where temporal ordering is fundamental to causal reasoning about bug introduction. The approach operates in two phases: (1) constructing a TKG that encodes commits with temporal and structural relationships, expanding the search space by traversing file history backward from two reference points (blame commits and the BFC); and (2) leveraging an LLM agent to navigate the graph using specialized tools for candidate exploration and causal analysis.

    Evaluation on three datasets shows that \appname achieves F1-scores of 0.47 to 0.79, with statistically significant F1 improvements over the state-of-the-art by up to 34\%. Our ablation study confirms that both components and the context expansion strategy each contribute: the TKG and agent form an exploration-exploitation synergy, while context expansion further unlocks ancestor BIC discovery, yielding a net gain of 60 true positives across datasets. A sensitivity analysis across five open-weight LLMs reveals that effective TKG navigation requires sufficiently capable models, and that the TKG architecture uniquely amplifies stronger LLMs, further widening the advantage over the state-of-the-art. By transforming BIC identification into a graph search problem, we open a new research direction for temporal and causal reasoning in software evolution analysis.
\end{abstract}




\received{20 February 2007}
\received[revised]{12 March 2009}
\received[accepted]{5 June 2009}

\maketitle

\section{Introduction}

% research context
Identifying the commit that introduced a bug, known as the Bug-Inducing Commit (BIC), is a foundational task in software maintenance. Accurate BIC identification enables understanding when and how bugs are introduced~\cite{DBLP:journals/tse/KameiSAHMSU13,DBLP:conf/msr/AsaduzzamanBRS12,DBLP:conf/scam/BavotaCLPOS12,DBLP:conf/msr/KimW06}, supports defect prediction models~\cite{DBLP:conf/icse/HataMK12,DBLP:journals/tse/YanXFHLL22,DBLP:journals/tse/FanXCLHL21,DBLP:conf/icse/TanTDM15,DBLP:journals/jss/PascarellaPB19}, and guides history-aware automated program repair systems~\cite{DBLP:journals/corr/abs-2511-01047} toward effective patches by providing precise historical context. For two decades, the dominant BIC identification approach has been the SZZ algorithm~\cite{DBLP:conf/msr/SliwerskiZZ05}, recognized with the 2026 ACM SIGSOFT Impact Paper Award, and its variants~\cite{DBLP:conf/kbse/KimZPW06,DBLP:conf/wcre/NetoCK18,DBLP:conf/kbse/TangBXH23,DBLP:journals/pacmse/TangLLYB25,DBLP:journals/tse/CostaMSKCH17,DBLP:journals/smr/DaviesRW14}, which trace deleted (and modified) code in a bug-fixing commit (BFC) via \texttt{git blame} to identify commits that last modified those lines. Presumably, one or more of those commits are the BIC.

% research gap
While SZZ operates on an implicit assumption that bug-inducing and bug-fixing commits modify the same code elements~\cite{DBLP:conf/sigsoft/WenWLTXCS19}, this assumption does not always hold: prior work has shown that over 13\% of BICs share no files with their corresponding bug-fixing commits, and 17\% of bug fixes contain only added lines where blame cannot be applied~\cite{DBLP:journals/tse/LyuKWLL24}. As the \texttt{git blame} mechanism constrains the search space to commits that directly touched the fixed lines, it misses BICs introduced through earlier code that was subsequently refactored~\cite{DBLP:conf/wcre/NetoCK18,DBLP:conf/esem/NetoCK19}, or bugs in related code where the fix addresses the issue by adding new functionality. Despite these fundamental limitations, prior work has focused primarily on improving precision of the git blame-based approach through filtering heuristics~\cite{DBLP:conf/kbse/KimZPW06, DBLP:journals/tse/CostaMSKCH17}, refactoring detection~\cite{DBLP:conf/wcre/NetoCK18,DBLP:conf/esem/NetoCK19}, deep learning~\cite{DBLP:conf/kbse/TangBXH23}, semantic analysis~\cite{DBLP:journals/pacmse/ChenZCWC25,DBLP:journals/tse/TangNHB24}, and LLM-based reasoning~\cite{DBLP:journals/pacmse/TangLLYB25}, leaving the recall problem unaddressed.

% research goal
\textbf{Our key insight is that BIC identification is not a ranking problem over a static list of blame candidates, but a search problem over a temporal graph of the commit history.}
A commit does not exist in isolation; each change builds upon previous code snapshots, forming a causal chain where later commits depend on earlier ones. When a blame commit merely reformatted code or made cosmetic changes, the actual bug may have been introduced by an earlier commit in this chain, or a later one. Identifying such BICs requires traversing the commit history: for example, following file modifications preceding or succeeding from a blame commit to find the commit that originally introduced the buggy logic. Temporal Knowledge Graphs (TKGs) are well-suited for this task because they explicitly encode the temporal dimension alongside structural relationships (shared files, functions, authors), enabling principled exploration of the commit history while respecting causality constraints.
While knowledge graphs have been successfully applied to various software engineering tasks such as automated program repair~\cite{DBLP:journals/corr/abs-2503-21710,DBLP:journals/corr/abs-2507-19942}, bug localization~\cite{DBLP:journals/pacmse/LiLZWLYLZ25}, and crash solution recommendation~\cite{DBLP:conf/sigsoft/DuLL0Y23}, they have not yet been applied to BIC identification.

% approach
To expand the search space beyond direct blame results, we present \appname, a novel approach that structures the BIC search space using a TKG and navigates it with an LLM agent. \appname operates in two phases:

\begin{itemize}
    \item \textbf{TKG Construction:} Given a bug-fixing commit, we build a graph encoding commits as nodes with temporal ordering and code structure connections (shared files and functions). Starting from two reference points, we expand the search space by traversing file history backward: from blame commits to capture earlier bug introductions, and from the BFC to capture commits that may have transformed correct code into buggy code.
    \item \textbf{Agentic BIC Search:} An LLM agent navigates the TKG using four specialized tools for candidate enumeration, structural traversal, property queries, and causal analysis. The agent reasons about which commits could have introduced the bug by examining code changes and their relationships.
\end{itemize}

% evaluation
We evaluate \appname on the same datasets used in prior work~\cite{DBLP:journals/pacmse/TangLLYB25}, comprising 2,102 validated BFC-BIC pairs. \appname achieves F1-scores of 0.47 to 0.79 across datasets, with F1 improvements from 16.0\% to 34.2\% over state-of-the-art. Our ablation study demonstrates that both components and the context expansion strategy each contribute: the TKG and agent are mutually essential (TKG-only introduces noise without intelligent filtering; agent-only is bounded by the blame-derived search space), and context expansion further unlocks ancestor BIC discovery. A sensitivity analysis across five open-weight LLMs reveals a capability threshold: only the two largest models (37B+ active parameters) consistently outperform the blame-fallback baseline, while smaller models make fewer tool calls per case, skipping the multi-step candidate verification that effective TKG navigation requires. A controlled comparison using a stronger LLM further shows that the TKG architecture amplifies the benefit, widening \appname's advantage over LLM4SZZ, while LLM4SZZ's blame-bounded search space limits its gains.
% A controlled comparison with the baselines using identical LLMs confirms that improvements stem from the TKG-guided paradigm rather than LLM selection.
% contributions
This paper makes the following contributions:
\begin{itemize}
    \item A preliminary study characterizing where ground truth BICs are located relative to blame commits, revealing that 28\% require commit history traversal and 14\% are blameless.
    \item \appname, the first approach to apply Temporal Knowledge Graphs to software evolution analysis, reframing BIC identification as a graph search problem and enabling identification of BICs beyond the blame-defined candidate set.
    \item Comprehensive evaluation demonstrating state-of-the-art performance, with ablation analysis validating the contribution of each component.
    \item An LLM sensitivity analysis across five open-weight models, establishing a capability threshold for effective TKG navigation and providing practical guidance for LLM selection.
\end{itemize}

\section{Background and Related Work}\label{sec:background}

\subsection{SZZ Algorithm and Its Variants}\label{subsec:szz-background}
% \textbf{SZZ Algorithm and Its Variants.}
The SZZ algorithm~\cite{DBLP:conf/msr/SliwerskiZZ05} pioneered automated BIC identification by tracing deleted or modified lines in bug-fixing commits via \texttt{git blame}. Over two decades, numerous variants have improved precision through annotation graphs and cosmetic change filtering~\cite{DBLP:conf/kbse/KimZPW06}, meta-change exclusion~\cite{DBLP:journals/tse/CostaMSKCH17}, refactoring detection~\cite{DBLP:conf/wcre/NetoCK18,DBLP:conf/esem/NetoCK19}, deep learning~\cite{DBLP:conf/kbse/TangBXH23}, and LLM-based reasoning~\cite{DBLP:journals/pacmse/TangLLYB25,11230098}. Most recently, LLM4SZZ~\cite{DBLP:journals/pacmse/TangLLYB25} leverages LLMs to rank blame candidates and expands context around modified lines for candidate collection. We provide detailed descriptions of our baselines in Section~\ref{subsec:baselines}. Despite these advances, all approaches share a fundamental limitation: they restrict the search space to commits returned by \texttt{git blame}, inherently missing BICs that require broader history traversal. \appname addresses this limitation by structuring the commit history as a temporal knowledge graph and navigating it through agent-guided exploration.

\subsection{Knowledge Graphs (KGs) in Software Engineering (SE)}\label{subsec:kg-background}
% \smallskip\noindent\textbf{Knowledge Graphs (KGs) in Software Engineering (SE).}
A KG is a graph-based data structure where nodes represent entities and edges encode relationships between them, enabling queries that traverse and reason over these connections \cite{hogan2021knowledge}. KGs have been increasingly applied to SE tasks. For automated program repair, repository-aware KGs bridge semantic gaps between issue descriptions and code patches~\cite{DBLP:journals/corr/abs-2503-21710}, and unified code KGs enable multilingual issue resolution~\cite{DBLP:journals/corr/abs-2507-19942}. For bug localization, injecting KG-derived knowledge into LLMs improves fault localization accuracy~\cite{DBLP:journals/pacmse/LiLZWLYLZ25}. Crash solution recommendation has been addressed by mining Stack Overflow into structured KGs~\cite{DBLP:conf/sigsoft/DuLL0Y23}. However, none of these works address BIC identification or model the temporal evolution of commits. Temporal knowledge graphs (TKGs) extend KGs with time-aware reasoning, enabling prediction over evolving relationships~\cite{DBLP:journals/corr/abs-2501-13956,DBLP:conf/icml/TrivediDWS17,DBLP:conf/wsdm/ParkLMCFD22}. To our knowledge, \appname is the first approach to apply TKGs to software evolution analysis. The temporal dimension is essential for BIC identification: a commit can only induce a bug if it precedes the fix, and understanding causal relationships requires reasoning about the sequence of code changes over time.

\subsection{LLM Agents for SE}\label{subsec:agents-background}
% \smallskip\noindent\textbf{LLM Agents for SE.}
LLM-based agents have emerged as powerful tools for complex SE tasks, marking a shift toward Agentic SE (SE 3.0) where agents work as AI teammates~\cite{DBLP:journals/corr/abs-2507-15003,DBLP:journals/corr/abs-2509-06216}. Recent approaches navigate codebases autonomously to resolve GitHub issues~\cite{DBLP:conf/nips/YangJWLYNP24,DBLP:conf/issta/0002RFR24}, using tool-augmented LLMs to search, read, and edit files. Simpler pipelines without persistent agent state have also proven effective~\cite{DBLP:journals/corr/abs-2407-01489}, general-purpose agent frameworks provide flexible foundations~\cite{DBLP:conf/iclr/0001LSXTZPSLSTL25}, and unified agents have been proposed to orchestrate multiple SE capabilities~\cite{DBLP:journals/corr/abs-2506-14683}. For bug fixing specifically, combining code KGs with agentic reasoning has shown promise~\cite{DBLP:journals/corr/abs-2409-00899}. An LLM-based framework has been proposed for BIC localization by integrating execution traces and commit artifacts~\cite{11230098}. However, this approach requires test execution for candidate filtering and does not model the temporal structure of commit history.
% These agents typically operate over code snapshots or file systems. 
In contrast, \appname operates over a TKG of commit history, using specialized tools designed for BIC search. This design enables reasoning about historical code evolution rather than static code states.

\section{Preliminary Study (RQ0): \rqzero}\label{sec:preliminary-study}

\subsection{Motivation}\label{subsec:rq0-motivation}

Existing SZZ techniques~\cite{DBLP:conf/msr/SliwerskiZZ05, DBLP:conf/kbse/KimZPW06, DBLP:journals/tse/CostaMSKCH17, DBLP:conf/wcre/NetoCK18} predominantly assume that bug-inducing commits (BICs) can be identified by applying \texttt{git blame} on the deleted or modified lines in the BFC. This is based on the common assumption that the bug fix location has to be one or more lines previously modified by the BIC \cite{DBLP:conf/msr/SliwerskiZZ05}. This assumption of course can be incorrect, for instance, when the commit identified by blame only modified whitespace characters and the real BIC is older than the former commit. In such cases, one could recursively apply blame to trace further back (illustrated as ``Blame Ancestor'' in Figure~\ref{fig:rq0-commits-category}); other times, the BIC actually could be more recent than the blame commit (``BFC Ancestor'' in Figure~\ref{fig:rq0-commits-category}), for instance when the BIC changes the value of a configuration option, which reveals a bug in a different part of the system. In such a situation, some other kind of code or commit relation should be used to navigate from the BFC to the BIC.

However, without knowing how often each scenario occurs and how deep the search must go, such extensions to the base SZZ techniques remain impractical. While prior work has examined file and statement overlap between BICs and bug-fixing commits (BFCs)~\cite{DBLP:conf/sigsoft/WenWLTXCS19}, the distribution of ground truth BICs across different positions in the commit history relative to blame results has not been systematically characterized. Understanding this distribution is essential for better guiding the search strategy across the commit history of a project. For instance, if one would find that BICs requiring backward traversal are common, how far back in history should one go to search the BIC? If BICs instead lie in between the blame commit and the BFC, how much closer to the BFC are they typically located?
% yes, fixed. Checked these papers they somehow touched, but didn't have ancestor check and such categorization
This preliminary study addresses this gap by categorizing where BICs appear with respect to blame commits, in order to assess the completeness of blame-based search and to determine whether broader history traversal is necessary.

% \hao{sounds like approach}We analyze 2,102 validated cases across DS\_LINUX, DS\_APACHE, and DS\_GITHUB to classify where BICs are located in the commit graph relative to the bug fix. This categorization provides a detailed breakdown of identification challenges. These findings directly motivate our approach design, specifically for the selection of BIC candidates nodes on a Temporal Knowledge Graph, which we evaluate in the subsequent research questions.

\subsection{Approach}\label{subsec:rq0-approach}

\subsubsection{Datasets}\label{subsubsec:datasets}

We use the same three developer-annotated datasets as LLM4SZZ~\cite{DBLP:journals/pacmse/TangLLYB25}, where BFC-BIC pairs are extracted from bug reports or commit messages authored by developers. These datasets provide higher ground truth quality than SZZ-generated datasets~\cite{DBLP:conf/esem/NetoCK19, DBLP:journals/smr/DaviesRW14}.
\textbf{DS\_LINUX}~\cite{DBLP:journals/tse/LyuKWLL24} is derived from the Linux kernel (written in C), where developers explicitly label bug-inducing commits in their fix messages. We use the exact same 1,500 pairs provided by LLM4SZZ~\cite{DBLP:journals/pacmse/TangLLYB25}.
% , who sampled from the original 76,046 pairs to ensure 95\% confidence level with margin of error under 5\%.
% \hao{I'm confused here, do we sample ourself or reuse the one sampled by this prior work? I do not think 95\% with 5\% lead to 1,500 pairs}fixed, I double checked their paper in page 12, yes.
\textbf{DS\_GITHUB}~\cite{DBLP:conf/icse/RosaPSTBLO21} spans 295 open-source repositories with developer-annotated BIC information mined from commit messages. This dataset contains 361 pairs, which we further split into DS\_GITHUB-c (C language, 286 pairs) and DS\_GITHUB-j (Java, 75 pairs) for language-specific analysis.
\textbf{DS\_APACHE}~\cite{DBLP:conf/sigsoft/WenWLTXCS19} is constructed from five Apache Software Foundation Java projects, with BIC annotations extracted from issue tracking systems and commit messages. This dataset contains 241 validated pairs.
In total, our evaluation covers 2,102 BFC-BIC pairs with 2,272 ground truth BICs, as some fixes address bugs introduced by multiple commits.

\subsubsection{BIC Categorization}\label{subsubsec:bic-categorization}

\begin{figure}[t]
    \centering
    \includegraphics[width=0.85\textwidth]{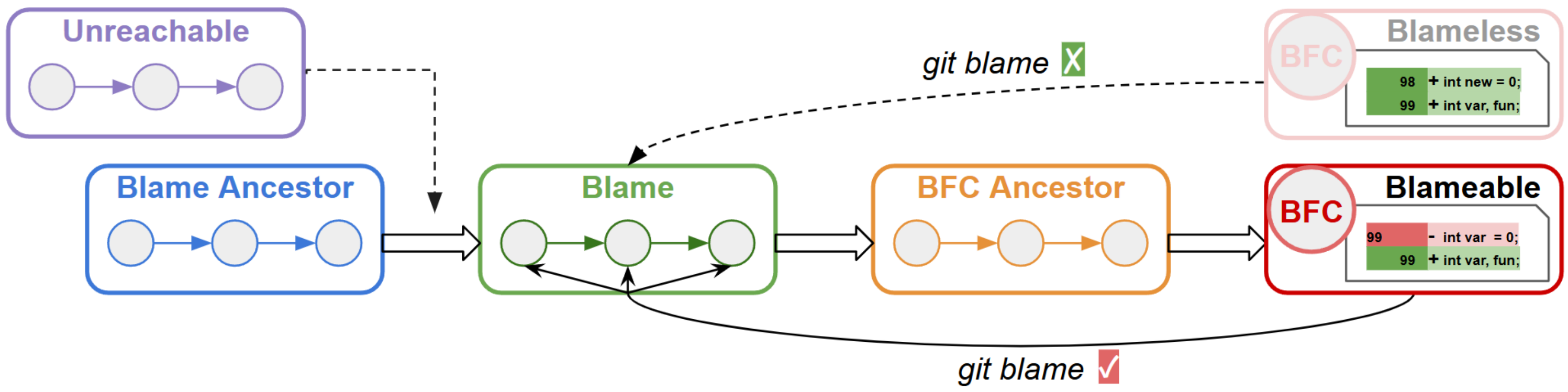}
    \caption{Categorization of bug-inducing commits (BICs) relative to blame commits. The Blame category (green) contains BICs that appear in blame commits; the Blame Ancestor category (blue) requires backward search from blame; the BFC Ancestor category (orange) requires backward search from the BFC toward blame.}
    \label{fig:rq0-commits-category}
\end{figure}

To characterize the search space, we execute \texttt{git blame} on the modified lines of each BFC and automatically classify the ground truth BIC based on its reachability from these blame results. We define five mutually exclusive categories (see Figure~\ref{fig:rq0-commits-category}). \textbf{Blame} includes BICs that appear directly in the blame set. \textbf{Blame Ancestor} includes BICs that are ancestors of blame commits and require backward traversal. \textbf{BFC Ancestor} includes BICs that are descendants of blame commits but precede the BFC, requiring traversal backward from the BFC.

We also account for cases where standard traversal fails. \textbf{Blameless} captures cases where the BFC contains no deleted or modified lines (i.e., only new code was added and nothing to \texttt{git blame}), rendering standard SZZ approaches inapplicable. Finally, \textbf{Unreachable} refers to BICs on parallel branches not reachable via backward traversal from blame commits. For example, suppose the BIC is on branch A while the blame commit is on branch B; after both branches merge, the buggy code is present, but \texttt{git blame} points to the branch B commit that last modified the same lines, making the true BIC on branch A unreachable through backward traversal. This classification determines the specific graph traversal capabilities required to identify the root cause.

% \hao{Why do we need file-history? This needs to be clarified and explained more clearly, for example, we try different ways (file history is one of them) to understand which one can reduce the search space. We may also need to report results based solely on commit distance, then compare them with file-history. In particular, we should report the finding that searching via file-history is faster than others in RQ0 to motivate our methodology.}Done. Added comparison. 
% \hao{mention the motivation in the beginning, e.g., `to determine whether broader history traversal is necessary'. Then, explain the approach for each category: 1. for Blame, we study the number of candidates (not sure about the term), 2. for Ancestor cases, we measure via file-history (explain the reason), 3. for blameless cases, we measure via fallback strategy.}fixed
To determine whether broader history traversal is necessary and to quantify the search space, we analyze each BIC category with a tailored measurement strategy. For the Blame category, we study the number of blame commits per case to understand candidate complexity. For the Blame Ancestor and BFC Ancestor categories, we measure the file-history depth required to reach the BIC. Given the scale of commit graphs (e.g., Linux has over 1 million commits), we filter candidates to those that modified exactly the same files as the BFC, as file co-modification is a necessary condition for code-level bug introduction. This file-scoped traversal is far more tractable than full DAG distance (median 13,786 commits across our raw datasets vs. median 2 commits for file-history) and focuses on commits most likely to be relevant. For Blameless cases where no deleted lines exist, we analyze a fallback strategy that applies \texttt{git blame} to the context lines surrounding each addition (2 lines before and after each added hunk~\cite{DBLP:journals/pacmse/TangLLYB25}, i.e., contiguous block of added lines), under the assumption that the BIC likely modified code adjacent to where the fix adds new logic.

% We categorize each ground truth BIC based on its position relative to blame commits of the BFC, establishing five mutually exclusive categories to guide search strategy design. Specifically, for each BFC, we execute \texttt{git blame} on the modified lines (i.e., deleted lines\hao{why modified lines are always deleted lines here? I got it after reading the full text}) to identify blame commits, which are the commits that last modified those lines before the fix. We then classify each ground truth BIC into one of five categories based on its position in the commit history relative to these blame commits.

\subsection{Results}\label{subsec:rq0-results}

\textbf{57.2\% of BICs are directly identifiable via git blame, while 28.0\% require extended commit history traversal (Blame Ancestor or BFC Ancestor).} Table~\ref{tab:rq0-bic-distribution} presents the category distribution across three datasets. The Blame category dominates with 57.2\% of all BICs, confirming that git blame provides a strong baseline for BIC identification. However, a substantial portion of BICs require traversing the commit graph beyond blame commits: 10.3\% are Blame Ancestors (requiring backward traversal from blame commits) and 17.7\% are BFC Ancestors (requiring backward traversal from the BFC). Notably, 14.1\% of cases are Blameless where traditional blame-based approaches cannot provide any candidates. These findings reveal that any approach relying solely on direct blame commits will miss nearly half of all BICs. Even for the majority Blame category, correctly identifying the true BIC among multiple blame candidates remains challenging.

\begin{table}[t]
    \centering
    \caption{Distribution of ground truth BIC categories across datasets.}
    \label{tab:rq0-bic-distribution}
    \footnotesize
    \begin{tabular}{l rr rrrrr}
        \toprule
        Dataset    & Cases & BICs  & Blame          & Blame Anc.   & BFC Anc.     & Blameless    & Unreachable \\
        \midrule
        DS\_LINUX  & 1,500 & 1,562 & 901 (57.7\%)   & 170 (10.9\%) & 219 (14.0\%) & 268 (17.2\%) & 4 (0.3\%)   \\
        DS\_GITHUB & 361   & 357   & 234 (65.5\%)   & 17 (4.8\%)   & 67 (18.8\%)  & 37 (10.4\%)  & 2 (0.6\%)   \\
        DS\_APACHE & 241   & 353   & 164 (46.5\%)   & 48 (13.6\%)  & 116 (32.9\%) & 15 (4.2\%)   & 10 (2.8\%)  \\
        \midrule
        Total      & 2,102 & 2,272 & 1,299 (57.2\%) & 235 (10.3\%) & 402 (17.7\%) & 320 (14.1\%) & 16 (0.7\%)  \\
        \bottomrule
    \end{tabular}
\end{table}

\begin{figure}[t]
    \centering
    \begin{subfigure}[b]{0.32\textwidth}
        \includegraphics[width=\textwidth]{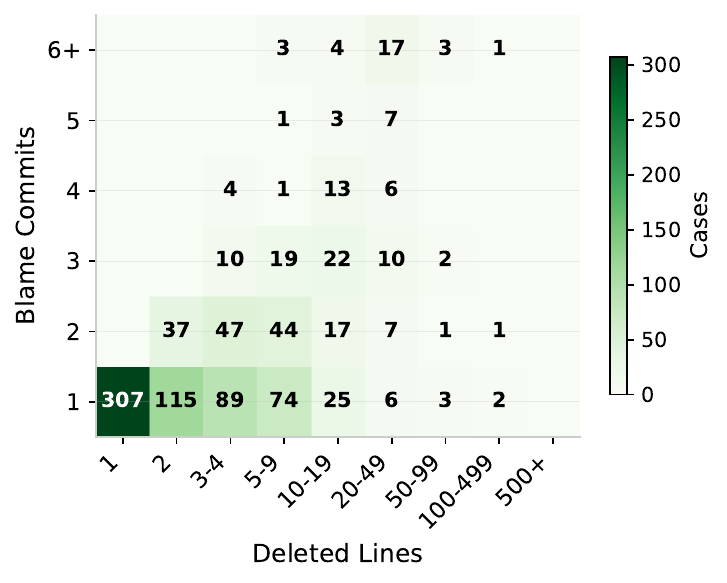}
        \caption{DS\_LINUX (n=901)}
        \label{fig:heatmap-linux}
    \end{subfigure}
    \hfill
    \begin{subfigure}[b]{0.32\textwidth}
        \includegraphics[width=\textwidth]{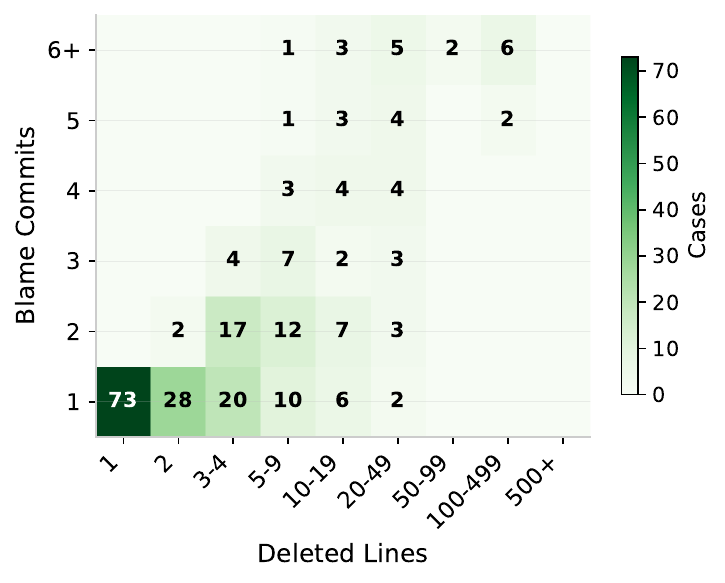}
        \caption{DS\_GITHUB (n=234)}
        \label{fig:heatmap-github}
    \end{subfigure}
    \hfill
    \begin{subfigure}[b]{0.32\textwidth}
        \includegraphics[width=\textwidth]{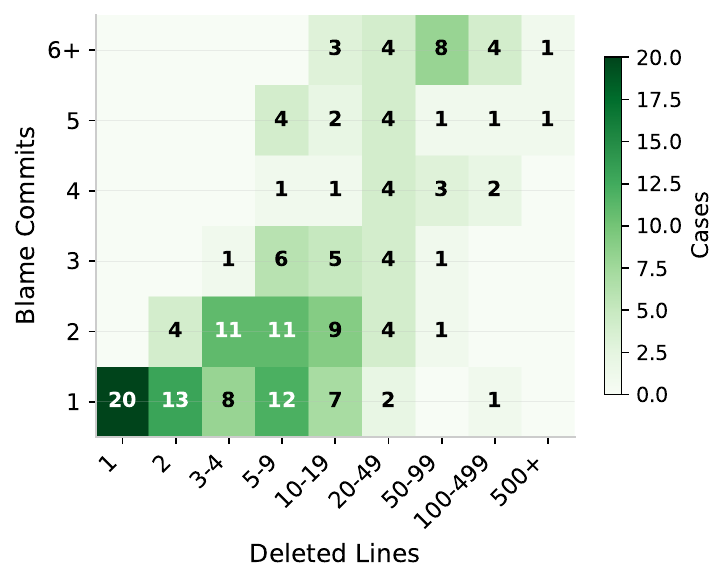}
        \caption{DS\_APACHE (n=164)}
        \label{fig:heatmap-apache}
    \end{subfigure}
    \caption{Distribution of blame complexity for Blame category cases. Each cell shows the number of cases with a given combination of deleted lines and blame commits.}
    \label{fig:complexity-heatmap}
    % \vspace{-0.5cm}
\end{figure}

\textbf{Even within the Blame category, identifying the correct BIC among multiple candidates is non-trivial.} Figure~\ref{fig:complexity-heatmap} shows the joint distribution of deleted lines and blame commits across datasets. The heatmap reveals distinct complexity patterns across datasets. DS\_LINUX and DS\_GITHUB concentrate in the low-complexity region (1 to 2 deleted lines with 1 blame commit): 422 cases (47\%) and 101 cases (43\%) respectively. In contrast, DS\_APACHE is more scattered, with only 33 cases (20\%) in this region but 49 cases (30\%) with $\geq$10 deleted lines and $\geq$3 blame commits.
% \hao{I'd suggest remove the definition of simple and complex, just report as it shown in the heatmap, e.g., 1 to 2 deleted lines with 1 blame commit}
Overall, 476 of 1,299 Blame category cases (36.6\%) involve multiple blame commits, requiring intelligent selection rather than simply returning all blame candidates.

\begin{figure}[t]
    \centering
    \begin{subfigure}[b]{0.35\textwidth}
        \includegraphics[width=\textwidth]{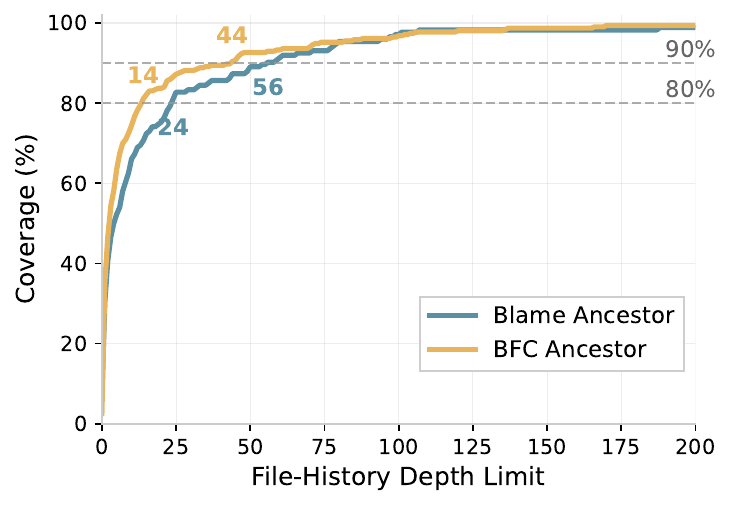}
        \caption{Blame Ancestor and BFC Ancestor.}
        \label{fig:rq0_file_depth_coverage}
    \end{subfigure}
    \hspace{1cm}
    \begin{subfigure}[b]{0.35\textwidth}
        \includegraphics[width=\textwidth]{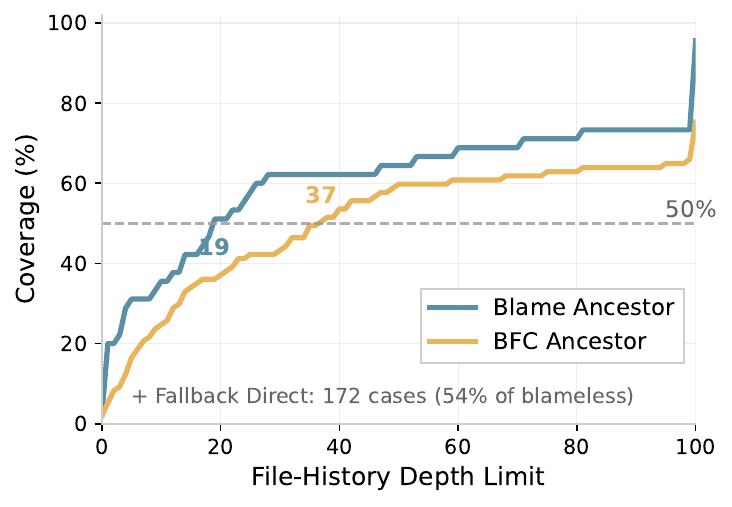}
        \caption{Blameless with fallback strategy.}
        \label{fig:rq0_blameless_coverage}
    \end{subfigure}
    \caption{File-history traversal depth required to reach BICs. (a) 28\% of cases require traversal beyond blame (Blame Ancestor and BFC Ancestor categories). (b) The 14.1\% of Blameless cases have no deleted lines.}
    \label{fig:search_depth_coverage_analysis}
\end{figure}

For the remaining 28\% of BICs that fall into the Blame Ancestor and BFC Ancestor categories, the challenge shifts from ranking candidates to determining how far to traverse the commit history. \textbf{File-history depth provides a tractable search space for BIC identification.} Figure~\ref{fig:search_depth_coverage_analysis} shows the percentage of BICs found as file-history traversal depth increases. For blameable cases (Figure~\ref{fig:rq0_file_depth_coverage}), traversing 24 ancestor commits covers 80\% of Blame Ancestor BICs (backward from blame), while 14 descendant commits cover 80\% of BFC Ancestor BICs (backward from BFC).

\textbf{Fallback strategy provides useful candidates for 54\% of Blameless cases.} For the 14.1\% of Blameless cases where git blame cannot be applied, we investigate whether blaming lines adjacent to additions could provide useful candidates. Figure~\ref{fig:rq0_blameless_coverage} shows that 54\% of Blameless BICs appear directly in these fallback blame commits. For the remaining cases, traversal depths of 19 (Blame Ancestor) to 37 (BFC Ancestor) commits achieve 50\% coverage.
% \hao{this sentence should be moved into approach}

\begin{summarybox}{Summary of RQ0}
    \small
    \begin{itemize}[leftmargin=*]
        \item While 57.2\% of BICs do appear directly in blame commits, 36.6\% of these Blame category cases involve multiple candidates requiring intelligent selection.
        \item 28.0\% of BICs require commit history traversal beyond blame results (Blame Ancestor or BFC Ancestor), and 14.1\% of cases are blameless (traditional blame-based approaches cannot apply).
    \end{itemize}
    \textbf{Implication:} Differentiating between these three categories is a promising direction for improving BIC identification. However, it requires intelligent reasoning to predict which category applies to a given case, as exhaustive graph traversal for all potential BIC candidates would be prohibitively expensive.
\end{summarybox}

\section{Methodology}\label{sec:methodology}
% \hao{I left some comments in the slides, need to make the TKG larger}fixed

\begin{figure}[t]
    \centering
    \includegraphics[width=0.95\textwidth]{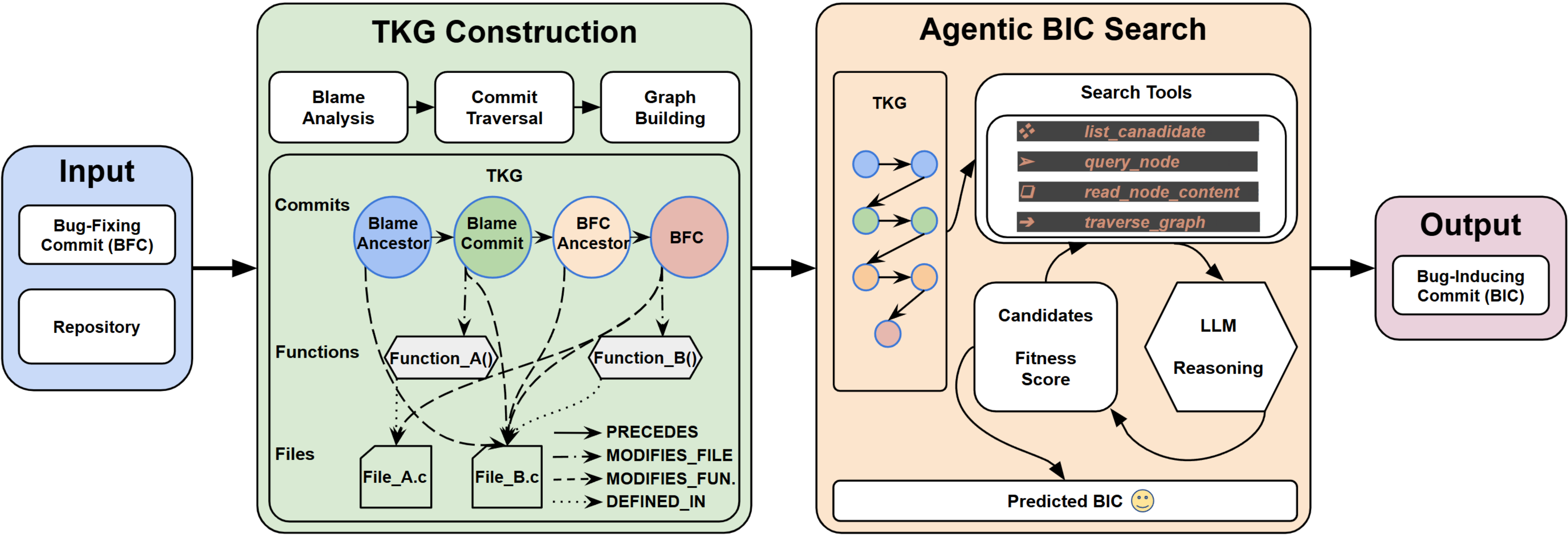}
    \caption{\appname architecture. Given a Bug-Fixing Commit (BFC) and its repository, TKG Construction identifies blame commits and expands the search space by traversing file history backward from two starting points (Blame Ancestor and BFC Ancestor). Agentic BIC Search then navigates the resulting graph using four specialized tools to identify the Bug-Inducing Commit (BIC).}
    \label{fig:methodology-overview}
\end{figure}

Based on the findings of the preliminary study, we propose a new BIC identification approach called \appname that goes beyond focusing only on commits of the Blame category, to which prior approaches are confined.
% Our preliminary study revealed that 28\% of BICs require traversing the commit history beyond blame results, and 14\% of cases are blameless where traditional blame cannot provide any candidates. These findings motivate our two-phase architecture
\appname has two phases as shown in Figure~\ref{fig:methodology-overview}: (1) constructing a Temporal Knowledge Graph (TKG) that expands the candidate set beyond direct blame commits to include ancestor commits, and (2) leveraging an LLM agent equipped with specialized tools to navigate and reason over this graph, allowing to identify other commits as the BIC, for instance, a BFC Ancestor or even a Blameless commit.

% Figure~\ref{fig:methodology-overview} illustrates the \appname pipeline. 

Given a bug-fixing commit (BFC) and its repository, we first construct a TKG through blame analysis and commit history traversal in two directions: backward from blame commits (Blame Ancestor) and backward from the BFC itself (BFC Ancestor). The TKG encodes commits as nodes with temporal ordering and code structure connections in the form of shared files and functions. An LLM agent then navigates this graph using specialized tools that support candidate enumeration, commit property inspection, causal analysis, and structural relationship exploration. The TKG should be as complete as possible to ensure the true BIC is reachable, while the agent must be as precise as possible to avoid overwhelming false positives from simply returning all graph nodes as candidates.

\subsection{Temporal Knowledge Graph Construction}\label{subsec:tkg-construction}

The TKG construction phase transforms raw repository data into a structured search space that captures both temporal commit relationships and code entity connections.

\subsubsection{Blame Analysis}
For each BFC, we extract the set of modified files and execute \texttt{git blame} on deleted and modified lines to identify blame commits. These are the commits that last touched the lines removed or changed by the fix, serving as initial BIC candidates. To improve ancestor coverage, we expand the blame set by including the deleted or modified line itself plus up to two existing context lines on each side before executing \texttt{git blame}. These surrounding lines can point to earlier commits that introduced nearby buggy logic, even when the directly blamed commit only modified the exact fixed line later. This heuristic is inspired by LLM4SZZ's context-enhanced strategy~\cite{DBLP:journals/pacmse/TangLLYB25}, yet the key distinction is that \appname uses the expanded blame as seeds for TKG construction and ancestor traversal rather than as the final candidate set. For blameless cases where the fix only adds code, we additionally apply a fallback that blames the context lines surrounding additions (2 lines before and 2 lines after each added hunk) to obtain initial commit candidates.

\subsubsection{Commit Traversal}
Based on the preliminary study findings (Section~\ref{sec:preliminary-study}) that 28\% of BICs require traversal beyond direct blame commits, we expand the search space through file-history traversal in two directions:

\begin{enumerate}[leftmargin=*]
    \item \textit{Blame Ancestor traversal}: Starting from each blame commit, we walk backward through the file history (using \texttt{git log --follow}) to collect commits that previously modified the same files. These candidates may have introduced bugs that were later touched but not fixed by the blamed commit.
    \item \textit{BFC Ancestor traversal}: Starting from the BFC, we walk backward through file history toward the blame commits. This captures commits that occurred after the blamed code was introduced but before the fix, which may have transformed correct code into buggy code.
\end{enumerate}
Depth limits (how many ancestor commits to traverse in each direction) and candidate caps (maximum number of commits to include in the TKG) are configurable parameters that balance coverage against computational cost; we discuss specific settings in Section~\ref{subsec:implementation}.

\subsubsection{Graph Building}
The graph contains three node types representing different aspects of the commit landscape:

\begin{enumerate}[leftmargin=*]
    \item \textit{Commit} nodes represent commits, storing commit messages, timestamps, modified files, extracted functions, and code diffs. Each commit is labeled with its type: \texttt{bfc}, \texttt{blame}, \texttt{blame\_ancestor}, or \texttt{bfc\_ancestor}. The BFC node additionally stores \textit{blame statistics}: the total number of blame commits, whether a single commit is responsible for all blamed lines, and whether a dominant commit accounts for the majority of blamed lines. These statistics enable fast-path decisions during agent search.
    \item \textit{File} nodes represent source files modified by commits.
    \item \textit{Function} nodes represent functions in which at least one line was changed by the union of identified commit nodes. We extract function names by parsing the hunk headers (i.e., the \texttt{@@} lines in \texttt{git diff} output that mark each contiguous block of changes and include the enclosing function name as context). Each function is associated with its containing file.
\end{enumerate}
Edges encode both temporal and structural relationships:

\begin{enumerate}[leftmargin=*]
    \item \texttt{PRECEDES} edges connect each commit to the next one by commit date within the TKG, regardless of branch structure or DAG parent-child relationships. This date-based linear ordering ignores the full DAG structure, but suffices for causality reasoning since a commit can only induce a bug if it precedes the fix in time.
    \item \texttt{MODIFIES\_FILE} and \texttt{MODIFIES\_FUNCTION} edges connect commits to the code entities they changed, enabling structural queries to find commits that touched the same code.
    \item \texttt{DEFINED\_IN} edges link functions to their containing files, enabling hierarchical code structure traversal.
\end{enumerate}

This schema enables the agent to reason about code-level relationships, identifying candidates that share modified functions or files with the BFC. Unlike prior works on knowledge graphs in software engineering (e.g., for automated program repair) \cite{DBLP:journals/corr/abs-2503-21710}, our approach explicitly leverages the temporal dimension of software evolution, constructing a temporal knowledge graph that preserves commit ordering and enables chronological reasoning about bug introduction.

\subsection{Agentic BIC Search}\label{subsec:search-agent}

As shown in Figure~\ref{fig:methodology-overview}, the Agentic BIC Search phase navigates the TKG to identify the most likely BIC through structural reasoning and causality verification. The search agent receives context about the BFC (modified files, extracted functions, commit message) and produces a prediction identifying the BIC commit with a reasoning chain.

\subsubsection{Design Rationale}
Our agentic loop design draws inspiration from recent advances in tool-augmented LLM agents for software engineering tasks~\cite{DBLP:conf/nips/YangJWLYNP24,DBLP:conf/issta/0002RFR24,DBLP:journals/corr/abs-2407-01489}. These agents demonstrate that LLMs can effectively navigate complex search spaces when provided with specialized tools and structured reasoning. We adapt this paradigm to BIC identification, where the agent must reason about causal relationships between code changes across time. Unlike agents for issue resolution that operate on static code snapshots, our agent operates on a temporal knowledge graph, using tools designed for graph traversal and temporal reasoning. The key insight is that an LLM can perform nuanced semantic analysis (e.g., distinguishing a refactoring from a bug introduction) that heuristic-based SZZ variants cannot, while the TKG provides the structured search space that enables efficient exploration.

\subsubsection{Search Tools}
The agent operates on the TKG through four specialized tools:

\begin{enumerate}[leftmargin=*]
    \item \textit{Candidate Enumeration} (\texttt{list\_candidates}): The agent's mandatory first step. Returns the top-$K$ candidates sorted by fitness score (see Algorithm~\ref{alg:agent-loop}), following the standard SZZ assumption that commits directly modifying fixed lines are most likely BICs. The tool also provides blame statistics to help the agent assess case complexity.
    \item \textit{Structural Traversal} (\texttt{traverse\_graph}): Follows \texttt{MODIFIES\_FILE} and \texttt{MODIFIES\_FUNCTION} edges to discover commits that modified the same code entities. Results are ordered by commit type (blame commits first, then blame ancestors, then BFC ancestors), with function-level matches prioritized over file-level matches since they indicate more precise code co-location.
    \item \textit{Property Query} (\texttt{query\_node}): Accesses metadata of a specific commit node, including the commit message, timestamp, modified files, and any code overlap with other candidates.
    \item \textit{Causal Analysis} (\texttt{read\_node\_content}): Reads the code diff stored in a commit node to perform causal analysis between BFC and BIC candidates. The agent checks if lines added by a candidate match what the BFC fixes, validating whether the candidate actually introduced the bug.
\end{enumerate}

\begin{algorithm}[t]
    \caption{Agentic BIC Search}
    \label{alg:agent-loop}
    \footnotesize
    \begin{algorithmic}[1]
        \Require TKG, BFC, top-$K$ limit
        \Ensure BIC prediction with reasoning

        \State \textbf{// Candidate ranking and retrieval}
        \State $candidates \gets$ \Call{list\_candidates}{TKG} \Comment{Returns top-$K$ ranked by fitness}
        \Statex \hspace{2em} $fitness(c) = \begin{cases} 1.0 & \text{blame} \\ 0.6 & \text{blame\_ancestor} \\ 0.3 & \text{bfc\_ancestor} \end{cases}$

        \State \textbf{// Agentic exploration loop}
        \State $step \gets 0$; $diff\_reads \gets 0$; $decision \gets \emptyset$
        \While{$step < 50$ \textbf{and} $decision = \emptyset$}
        \State $action \gets$ \Call{LLM\_Reason}{$candidates$, output}
        \If{$action = $ \texttt{query\_node}$(sha)$}
        \State output $\gets$ \Call{query\_node}{TKG, $sha$}
        \ElsIf{$action = $ \texttt{read\_node\_content}$(sha)$ \textbf{and} $diff\_reads < 3$}
        \State output $\gets$ \Call{read\_node\_content}{TKG, $sha$}
        \State $diff\_reads \gets diff\_reads + 1$
        \ElsIf{$action = $ \texttt{traverse\_graph}$(sha)$}
        \State output $\gets$ \Call{traverse\_graph}{TKG, $sha$}
        \ElsIf{$action = $ decide$(sha, reason)$} \Comment{Agent decision}
        \State $decision \gets (sha, reason)$
        \EndIf
        \State $step \gets step + 1$
        \EndWhile
        \State \Return $decision$
    \end{algorithmic}
\end{algorithm}

\subsubsection{Search Strategy}
% \hao{need to explain more how to determine the score}\hao{why do we need this? I feel list\_candidates is enough. Difficult to defend without empirical evidence. Also, make the agent's decision strategy less generalizable}\hao{Why? Which evidence supports this rule? In RQ0, we just mentioned 28\% need to check ancestor cases. Maybe just give it high score instead of exit}\hao{why do we need this if list\_candidates already indicate it?}fixed: removed fast-paths, simplified to ordinal ranking grounded in RQ0 and prior work
Algorithm~\ref{alg:agent-loop} presents the agentic search loop. The agent begins by calling \texttt{list\_candidates} to obtain all candidate commits ranked by fitness score. This ranking prioritizes commit types based on our RQ0 findings: blame commits receive the highest score (1.0) since 57.2\% of BICs appear directly in blame results; blame ancestors receive 0.6 as they require one level of traversal; BFC ancestors receive 0.3 as they represent the broadest exploration. These scores serve as ordinal rankings to guide the agent's attention; the key property is the relative ordering, other values preserving this pattern would yield similar results.

The agent then iteratively investigates the ranked candidates using the available tools. At each step, the LLM receives the full ranked candidate list along with outputs accumulated from previous tool calls (the `output' variable in Algorithm~\ref{alg:agent-loop}), and may choose any candidate to investigate: querying its properties, reading its code diff to verify causal relationships, or traversing the graph to find related commits. In other words, the LLM could override the initial ranking returned by \texttt{list\_candidates} based on its reasoning process. The loop continues until the agent reaches a decision or exhausts the maximum step limit (50 steps). To control API costs, we empirically limit diff reads to 3 per case. The agent outputs a structured prediction containing the predicted BIC SHA and reasoning chain.

\section{Evaluation Setup}\label{sec:evaluation-setup}

We use the same three datasets (DS\_LINUX, DS\_APACHE, and DS\_GITHUB) described in Section~\ref{subsubsec:datasets}, comprising 2,102 validated BFC-BIC pairs with 2,272 ground truth BICs.

\subsection{Baselines}\label{subsec:baselines}

We compare \appname against seven SZZ variants and one LLM-based approach, adopting the same baselines as LLM4SZZ~\cite{DBLP:journals/pacmse/TangLLYB25} to enable direct comparison. Since we use identical datasets and evaluation methodology, we report baseline results from this prior work where available.\footnote{Baseline results for B-SZZ, AG-SZZ, MA-SZZ, R-SZZ, L-SZZ, RA-SZZ, and Neural-SZZ are from the LLM4SZZ paper~\cite{DBLP:journals/pacmse/TangLLYB25}.}

% \hao{Are these methods based on blame?}yes, added clarification
All SZZ baseline variants are solely based on \texttt{git blame} over deleted or modified lines in the BFC. They differ primarily in their filtering and ranking strategies:
\textbf{B-SZZ}~\cite{DBLP:conf/msr/SliwerskiZZ05} is the original SZZ algorithm that traces deleted lines via \texttt{git blame} and returns all blame commits as candidates.
\textbf{AG-SZZ}~\cite{DBLP:conf/kbse/KimZPW06} extends B-SZZ by filtering cosmetic changes (blank lines, comments) and using annotation graphs (which map line origins across revisions, even when lines move due to insertions or deletions) for more precise line tracking.
\textbf{MA-SZZ}~\cite{DBLP:journals/tse/CostaMSKCH17} further excludes meta-changes (branch merges, property changes) that do not affect program behavior.
\textbf{R-SZZ} and \textbf{L-SZZ}~\cite{DBLP:journals/smr/DaviesRW14} select a single BIC from AG-SZZ candidates: R-SZZ chooses the most recent commit, while L-SZZ chooses the commit with the most changed lines.
\textbf{RA-SZZ}~\cite{DBLP:conf/wcre/NetoCK18} filters refactoring operations using RefDiff~\cite{DBLP:conf/msr/SilvaV17} and Refactoring Miner~\cite{DBLP:conf/icse/TsantalisMEMD18}, but is limited to Java projects.
\textbf{Neural-SZZ}~\cite{DBLP:conf/kbse/TangBXH23} applies deep learning by embedding code changes with CodeBERT~\cite{DBLP:conf/emnlp/FengGTDFGS0LJZ20} and ranking them using a heterogeneous graph attention network. It is also limited to Java.

\textbf{LLM4SZZ}~\cite{DBLP:journals/pacmse/TangLLYB25} is the current state-of-the-art approach that leverages LLMs to identify BICs from blame candidates. It employs two strategies: rank-based identification that asks the LLM to rank buggy statements by their relevance to the root cause, and context-enhanced identification that provides additional context for direct BIC selection. LLM4SZZ expands context around modified lines (full function bodies or $\pm$2 surrounding lines) and blames these expanded lines as well for candidate collection. This enables LLM4SZZ to handle blameless cases and reach some BICs beyond the immediate deleted-line blame set. However, LLM4SZZ does not traverse commit history, which is the gap that \appname addresses through TKG-based exploration.
% \hao{Can these methods also detect ancestors?}no, added clarification that they are restricted to blame set

In summary, LLM4SZZ's context expansion extends blame beyond deleted lines but does not traverse commit history. Furthermore, only LLM4SZZ handles blameless cases; none of the other baselines can provide any candidates when the bug fix contains only additions.
% \hao{and LLM4SZZ is the only baseline consider blameless? we can also mention this}fixed

\subsection{Evaluation Metrics}\label{subsec:metrics}

Following LLM4SZZ, we adopt standard information retrieval metrics. Let $P$ denote the set of predicted BICs and $G$ denote the set of ground truth BICs for a given BFC.
\textbf{Precision} measures the fraction of predictions that match ground truth BICs, \textbf{Recall} measures the fraction of ground truth BICs that are correctly identified, and \textbf{F1-score} is the harmonic mean of Precision and Recall, providing a balanced measure:

\begin{equation}
    \text{Precision} = \frac{|P \cap G|}{|P|},
    \ \text{Recall} = \frac{|P \cap G|}{|G|},
    \ \text{F1-score} = \frac{2 \times \text{Precision} \times \text{Recall}}{\text{Precision} + \text{Recall}}
\end{equation}

For BFCs with multiple ground truth BICs, a prediction is considered correct if it matches any valid BIC in $G$. Following prior work~\cite{DBLP:journals/pacmse/TangLLYB25}, each approach returns a single predicted BIC per case.

\subsection{Implementation}\label{subsec:implementation}

\appname is implemented in Python 3.12 using the following components:

\begin{enumerate}[leftmargin=*]
    \item \textit{LLM Backend.} We use DeepSeek-V3.2~\cite{DBLP:journals/corr/abs-2512-02556} as the agent's reasoning engine, accessed via the official API. The LLM operates with temperature 0 for reproducibility.

    \item \textit{Knowledge Graph.} The TKG is stored in Neo4j\footnote{\url{https://neo4j.com/}} using the Graphiti framework~\cite{DBLP:journals/corr/abs-2501-13956} for temporal knowledge graph management. Graphiti provides native support for temporal entity and edge management, enabling efficient retrieval of commits and their relationships while maintaining chronological ordering. We design the domain-specific schema (commit, file, and function nodes; temporal and structural edges) and tool APIs on top of this general-purpose framework.

    \item \textit{TKG Configuration.} For commit traversal, we set the maximum file-history depth to 100 commits in each direction (BFC ancestor and blame ancestor), with a hard cap of 200 total candidates per TKG. These generous upper bounds ensure high coverage (RQ0 shows that 80\% of ancestor BICs are reachable within depth 24); the actual number of commits collected depends on each file's history length. The agent evaluates the top-$K$=20 candidates ranked by fitness score; we analyze sensitivity to $K$ in Section~\ref{sec:threats}.

    \item \textit{Agent Framework.} We implement the search agent using pydantic-ai,\footnote{\url{https://ai.pydantic.dev/}} a Python agent framework that provides typed tool interfaces and structured output parsing. The agent receives the BFC context (commit message, modified files, extracted functions) and iteratively calls tools until it reaches a decision or exhausts the step limit of 50.

    \item \textit{Computational Resources.} Experiments were conducted on a 48-core server with 1.5TB RAM. For parallel evaluation, each worker runs an isolated Neo4j instance in Docker, enabling concurrent processing across multiple cases.
\end{enumerate}

For the main comparison (Table~\ref{tab:rq1-results}) and controlled comparison (Table~\ref{tab:llm4szz-vs-agenticszz}), we follow the evaluation protocol of LLM4SZZ~\cite{DBLP:journals/pacmse/TangLLYB25}: both approaches are run three times independently, and we report metrics averaged across runs to reduce variance from LLM non-determinism. For statistical tests, we compare the best-performing run from each approach; we additionally test robustness by comparing our worst run against baselines' best. For cost analysis (Table~\ref{tab:cost-average}), we report the average processing time and cost per case across all three runs. For the ablation study and category breakdown analysis (RQ2), we report results from a single run, as our RQ1 results showed that performance overall is stable across runs (both for AgenticSZZ and LLM4SZZ).

\subsection{Data Leakage Prevention}\label{subsec:data-leakage}

Commit messages in real repositories can contain explicit references to the BIC, added by developers who manually identified the bug-inducing commit. This is particularly prevalent in the Linux kernel, which follows a structured commit message format including \texttt{Fixes:} tags that directly reference the bug-inducing commit (this convention is how the DS\_LINUX BIC dataset was constructed). Since SZZ aims to automate this manual identification process, we must prevent the agent from exploiting these ground truth hints during evaluation.

For this reason, we sanitize all commit messages during TKG construction by removing \texttt{Fixes:} lines, \texttt{Reverts:} tags, and replacing inline commit SHA references with placeholder tokens. This ensures the agent must reason about code relationships rather than exploiting metadata leakage.

\section{Evaluation Results}\label{sec:evaluation-results}

\subsection{RQ1: \rqone}\label{subsec:rq1}

\subsubsection{Motivation}\label{subsubsec:rq1-motivation}

% \hao{motivate why BIC is important and why we need TKG and agent, i.e., motivate the problem not the solution}fixed, focus on the limitations of existing approaches
Our preliminary study (RQ0) revealed that existing blame-based approaches face fundamental limitations: a significant portion of BICs require traversing commit history beyond direct blame results, and many cases involve multiple candidates requiring intelligent selection.
% First, while 57.2\% of BICs appear in blame commits, 33.4\% of these cases involve multiple candidates, requiring intelligent selection rather than simple heuristics. Second, 28.0\% of BICs lie outside the immediate blame set but remain reachable through commit history traversal (Blame Ancestor or BFC Ancestor categories). Third, 14.1\% of cases are blameless, where traditional blame provides no candidates and fallback strategies are needed. 
% \hao{no neeed to repeat the number, summarize the key insights instead.}fixed, just we summarize insights
Current state-of-the-art approaches like LLM4SZZ~\cite{DBLP:journals/pacmse/TangLLYB25} expand context around modified lines for blame candidate collection (Section~\ref{subsec:baselines}) but do not traverse commit history to systematically explore ancestor commits.
% This motivates our core research question: can a Temporal Knowledge Graph (TKG) combined with an LLM agent effectively navigate this expanded search space to identify BICs?

\subsubsection{Approach}\label{subsubsec:rq1-approach}

We evaluate \appname on the complete set of 2,102 bug-fixing cases across DS\_LINUX (1,500 cases), DS\_APACHE (241 cases), and DS\_GITHUB (361 cases). DS\_GITHUB is further split into DS\_GITHUB-c (C language, 286 cases) and DS\_GITHUB-j (Java, 75 cases) to enable language-specific analysis, as Neural-SZZ and RA-SZZ are Java-only baselines. For each case, \appname constructs a TKG through blame analysis and commit history traversal from two starting points, then employs an LLM agent to navigate the graph using the four specialized tools described in Section~\ref{subsec:search-agent}. We compare against eight baselines (Section~\ref{subsec:baselines}), including six SZZ variants, Neural-SZZ, and LLM4SZZ.

For the controlled comparison (Table~\ref{tab:llm4szz-vs-agenticszz}), both approaches use DeepSeek-V3.2, isolating the architectural contribution from LLM selection.

All approaches are evaluated using Precision, Recall, and F1-score as defined in Section~\ref{subsec:metrics}.
We use McNemar's test for paired binary outcomes (correct/incorrect per case) and report Cohen's $g$ as effect size, where $g \geq 0.15$ indicates a medium effect and $g \geq 0.25$ indicates a large effect~\cite{cohen2013statistical}.
% \hao{mention deepseek for LLM4SZZ here}fixed
% \hao{need one sentence to explain MCNemar and effect size}fixed

\subsubsection{Results}\label{subsubsec:rq1-results}

% Merged RQ1 results table (merged Table 3 and 4 per Hao's suggestion to save space)
\begin{table*}[t]
    \centering
    \caption{\appname effectiveness across all datasets (averaged over three runs). Underlined: best baseline; ``-'': Java-only baselines not applicable to C projects.}
    \label{tab:rq1-results}
    \footnotesize
    \begin{tabular}{l rrr rrr rrr rrr}
        \toprule
        \multirow{2}{*}{\textbf{Method}} & \multicolumn{3}{c}{\textbf{DS\_LINUX}} & \multicolumn{3}{c}{\textbf{DS\_GITHUB-c}} & \multicolumn{3}{c}{\textbf{DS\_GITHUB-j}} & \multicolumn{3}{c}{\textbf{DS\_APACHE}}                                                                                                                                                                 \\
        \cmidrule(lr){2-4} \cmidrule(lr){5-7} \cmidrule(lr){8-10} \cmidrule(lr){11-13}
                                         & Prec.                                  & Rec.                                      & F1                                        & Prec.                                   & Rec.              & F1                & Prec.             & Rec.              & F1                & Prec.             & Rec.              & F1                \\
        \midrule
        B-SZZ                            & 0.452                                  & \underline{0.578}                         & 0.507                                     & 0.361                                   & \underline{0.656} & 0.466             & 0.285             & \underline{0.680} & 0.401             & 0.251             & \underline{0.435} & 0.318             \\
        AG-SZZ                           & 0.448                                  & 0.553                                     & 0.495                                     & 0.410                                   & 0.592             & 0.484             & 0.421             & 0.533             & 0.470             & 0.328             & 0.310             & 0.318             \\
        MA-SZZ                           & 0.421                                  & 0.538                                     & 0.472                                     & 0.335                                   & 0.624             & 0.436             & 0.239             & 0.560             & 0.335             & 0.307             & 0.345             & 0.329             \\
        R-SZZ                            & 0.583                                  & 0.448                                     & 0.507                                     & 0.671                                   & 0.582             & 0.620             & 0.538             & 0.467             & 0.500             & 0.497             & 0.288             & 0.364             \\
        L-SZZ                            & 0.560                                  & 0.430                                     & 0.486                                     & 0.486                                   & 0.422             & 0.452             & 0.492             & 0.427             & 0.457             & 0.366             & 0.211             & 0.267             \\
        RA-SZZ                           & -                                      & -                                         & -                                         & -                                       & -                 & -                 & 0.337             & 0.440             & 0.382             & 0.264             & 0.325             & 0.293             \\
        Neural-SZZ                       & -                                      & -                                         & -                                         & -                                       & -                 & -                 & 0.556             & 0.486             & 0.520             & 0.563             & 0.364             & 0.442             \\
        LLM4SZZ                          & \underline{0.628}                      & 0.552                                     & \underline{0.588}                         & \underline{0.687}                       & 0.641             & \underline{0.663} & \underline{0.607} & 0.569             & \underline{0.587} & \underline{0.610} & 0.398             & \underline{0.482} \\
        \midrule
        \textbf{\appname}                & \textbf{0.696}                         & \textbf{0.669}                            & \textbf{0.682}                            & \textbf{0.780}                          & \textbf{0.777}    & \textbf{0.779}    & \textbf{0.788}    & \textbf{0.788}    & \textbf{0.788}    & \textbf{0.575}    & \textbf{0.394}    & \textbf{0.468}    \\
        \bottomrule
    \end{tabular}%

\end{table*}

% \hao{avoid repeating the numbers in the tables, report the best and provide some insights about why we are better than SOTA, if we have nothing to say, then just one paragraph by saying we have the best performance. But need to explain why (insights) not the best in DS\_APACHE.}fixed
\textbf{\appname achieves the best F1-score on three of four datasets.} Table~\ref{tab:rq1-results} presents results across all four dataset splits. \appname outperforms all baselines including LLM4SZZ, the previous state-of-the-art, with improvements of 17.5\% on DS\_GITHUB-c, 16.0\% on DS\_LINUX, and 34.2\% on DS\_GITHUB-j; DS\_APACHE shows slightly lower but statistically comparable performance (0.468 vs. 0.482, $p=1.000$). Using the best-performing run from each approach, the paired test over all matched cases across the RQ1 datasets shows that \appname significantly outperforms LLM4SZZ (McNemar's test, $p<0.001$, Cohen's $g=0.23$, medium effect). These run-level statistical results complement Table~\ref{tab:rq1-results}, which reports metrics averaged over three runs. As a robustness check, \appname's worst-performing run still outperforms LLM4SZZ's best run over all matched cases ($p<0.001$, $g=0.21$), showing that the overall improvement is not driven by a single favorable run.
% \hao{this is the threshold for strong??}\hao{medium?}fixed
Among baselines, traditional SZZ variants like B-SZZ achieve higher recall by returning multiple candidates, but suffer from low precision; conversely, heuristics like R-SZZ improve precision but sacrifice recall. \appname achieves both the highest recall and highest precision by expanding the search space through the TKG while using intelligent agent reasoning to filter candidates.

\textbf{The recall improvements validate our RQ0 findings about BIC distribution.} \appname achieves higher recall than LLM4SZZ on DS\_LINUX, DS\_GITHUB-c, and DS\_GITHUB-j, with the largest gains on DS\_GITHUB-j (+0.219) and DS\_LINUX (+0.117), while DS\_APACHE shows marginally lower recall (0.394 vs. 0.398). These improvements reflect our RQ0 finding that 28\% of BICs require exploration beyond immediate blame commits. While LLM4SZZ's context expansion can reach some of these BICs, \appname's TKG-based history traversal provides a more systematic approach to exploring ancestor commits, as the BIC category analysis in RQ2 confirms. The simultaneous precision improvements indicate that the agent's reasoning effectively filters false positives introduced by search space expansion.

\textbf{DS\_APACHE presents unique challenges that limit improvement.} While \appname falls slightly below LLM4SZZ on DS\_APACHE in terms of F1 (0.468 vs. 0.482), the difference is not statistically significant ($p=1.000$). As shown in Figure~\ref{fig:complexity-heatmap}, DS\_APACHE has the highest blame complexity: 30\% of cases involve $\geq$10 deleted lines with $\geq$3 blame commits, whereas DS\_LINUX and DS\_GITHUB concentrate on the low-complexity region (47\% and 43\% respectively have 1 to 2 deleted lines with only 1 blame commit). This complexity increases the difficulty of identifying the correct BIC among many candidates. Additionally, DS\_APACHE projects span diverse Apache Software Foundation repositories with varying coding conventions, making it harder for the agent to reason about code relationships consistently.

\textbf{These observed performance improvements come with moderate processing overhead.} Table~\ref{tab:cost-average} presents the average processing time per case. \appname requires 42 to 80 seconds per case, compared to 20 to 30 seconds for LLM4SZZ~\cite{DBLP:journals/pacmse/TangLLYB25}. The additional time is primarily due to TKG construction (12 to 49 seconds depending on repository size), while agent reasoning time (approximately 30 seconds) is consistent across datasets. DS\_LINUX has the longest TKG construction time (49 seconds) because git operations are slower on large repositories. Compared to RA-SZZ, which requires 78 seconds per case for refactoring detection~\cite{DBLP:journals/pacmse/TangLLYB25}, \appname averages 58 seconds per case while achieving substantially higher accuracy.

\begin{table}[t]
    \centering
    \caption{\appname average processing time and cost per case. Cost is calculated based on DeepSeek-V3.2 API pricing as of December 2025.\protect\footnotemark}
    \label{tab:cost-average}
    \footnotesize
    \begin{tabular}{l rrr rr r}
        \toprule
        \textbf{Dataset} & \textbf{Total (s)} & \textbf{TKG (s)} & \textbf{Agent (s)} & \textbf{\# Input Tokens} & \textbf{\# Output Tokens} & \textbf{Cost (\$)} \\
        \midrule
        DS\_LINUX        & 79.8               & 48.9             & 30.2               & 45,407                   & 615                       & 0.004              \\
        DS\_GITHUB       & 50.9               & 20.0             & 30.6               & 43,463                   & 643                       & 0.004              \\
        DS\_APACHE       & 42.3               & 11.9             & 29.9               & 55,279                   & 652                       & 0.005              \\
        \bottomrule
    \end{tabular}
\end{table}
\footnotetext{\url{https://api-docs.deepseek.com/quick_start/pricing}}

% Table moved from Discussion per Hao's comment
\begin{table*}[t]
    \centering
    \caption{Controlled comparison: both approaches use DeepSeek-V3.2, averaged over three runs. $^{*}$Significant ($p<0.05$, McNemar's test).}
    \label{tab:llm4szz-vs-agenticszz}
    \footnotesize
    \begin{tabular}{l rrr rrr rrr rrr}
        \toprule
        \multirow{2}{*}{\textbf{Approach}} & \multicolumn{3}{c}{\textbf{DS\_LINUX}} & \multicolumn{3}{c}{\textbf{DS\_GITHUB}} & \multicolumn{3}{c}{\textbf{DS\_APACHE}} & \multicolumn{3}{c}{\textbf{Overall}}                                                                                                                                                     \\
        \cmidrule(lr){2-4} \cmidrule(lr){5-7} \cmidrule(lr){8-10} \cmidrule(lr){11-13}
                                           & Prec.                                  & Rec.                                    & F1                                      & Prec.                                & Rec.           & F1                   & Prec.          & Rec.           & F1             & Prec.          & Rec.           & F1                   \\
        \midrule
        LLM4SZZ                            & 0.595                                  & 0.584                                   & 0.590                                   & 0.611                                & 0.460          & 0.525                & 0.537          & \textbf{0.412} & 0.467          & 0.590          & 0.538          & 0.563                \\
        \appname                           & \textbf{0.696}                         & \textbf{0.669}                          & \textbf{0.682}$^{*}$                    & \textbf{0.782}                       & \textbf{0.780} & \textbf{0.781}$^{*}$ & \textbf{0.575} & 0.394          & \textbf{0.468} & \textbf{0.697} & \textbf{0.644} & \textbf{0.669}$^{*}$ \\
        \bottomrule
    \end{tabular}
\end{table*}

% \hao{deepseek + LLM4SZZ results can also be included here}moved here from Discussion
\textbf{Controlled comparison confirms that improvements mainly stem from the TKG-guided paradigm, not merely from recent advances in the underlying LLM.} Since \appname uses DeepSeek-V3.2 while LLM4SZZ employs GPT-4o-mini, one might question whether \appname's improvements are due to architectural contributions or simply a more capable LLM. To isolate the architectural impact, we conducted a controlled comparison (Table~\ref{tab:llm4szz-vs-agenticszz}) where both approaches use DeepSeek-V3.2. LLM4SZZ with DeepSeek achieves comparable F1-scores to the original GPT-4o-mini version (e.g., 0.590 vs 0.588 on DS\_LINUX), indicating that DeepSeek is at least as capable for this task. Comparing against this controlled baseline, \appname achieves an 18.8\% relative improvement in overall F1-score (0.669 vs 0.563), due to gains of 15.6\% on DS\_LINUX and 48.8\% on DS\_GITHUB. On DS\_APACHE, both approaches achieve comparable F1-scores (0.468 vs 0.467), consistent with the blame complexity challenges discussed above. The best-vs-best comparison is statistically significant (McNemar's test, $p<0.001$, Cohen's $g=0.27$, large effect); even our worst run outperforms their best ($p<0.001$, $g=0.26$), demonstrating that the TKG-guided search paradigm, not merely LLM capability or run variance, drives \appname's effectiveness.

\begin{summarybox}{Summary of RQ1}
    \small
    \begin{itemize}[leftmargin=*]
        \item \appname achieves the best F1-score on three of four dataset splits, with F1 improvements from 16.0\% to 34.2\% over LLM4SZZ; even \appname's worst run beats LLM4SZZ's best run over all matched cases ($p<0.001$).
        \item Controlled comparison confirms improvements stem from the TKG-guided paradigm, not LLM selection.
    \end{itemize}
    \textbf{Implication:} Reframing BIC identification as graph search over a TKG advances the state-of-the-art.
\end{summarybox}

\subsection{RQ2: \rqtwo}\label{subsec:rq2}

\subsubsection{Motivation}\label{subsubsec:rq2-motivation}

\appname integrates two core components: a Temporal Knowledge Graph (TKG) for search space construction and expansion, and an LLM agent for intelligent navigation. The design also expands each blame seed with up to two existing context lines before and after it, whose contribution we isolate as a third ablation dimension. Understanding each component's contribution is essential for validating our design decisions and guiding future research. We conduct an ablation study to answer: Is the TKG necessary, or can an LLM agent reason effectively over raw blame results alone? Does the fallback strategy for blameless cases contribute meaningfully? Does combining TKG and agent produce synergistic benefits beyond their individual contributions? Does context expansion consistently improve performance across datasets?

\subsubsection{Approach}\label{subsubsec:rq2-approach}

We evaluate six configurations that systematically enable or disable components:

\begin{enumerate}[leftmargin=*]
    \item \textit{Blame-only}: Returns the most recent blame commit as the predicted BIC, following traditional SZZ heuristics. No LLM reasoning, no TKG traversal, and no fallback for blameless cases (returns empty prediction). This represents the simplest heuristic baseline.

    \item \textit{Blame-fallback}: Extends Blame-only by applying the fallback strategy for blameless cases, blaming $\pm$2 lines around additions to obtain candidates. This isolates the contribution of handling BIC identification of blameless bug fixes.

    \item \textit{TKG-only}: Constructs the full TKG with commit history traversal from two starting points but replaces the LLM agent with deterministic selection. Unlike the full pipeline shown in Algorithm~\ref{alg:agent-loop} where the agent iteratively investigates candidates (reading diffs, traversing relationships, reasoning about causality), TKG-only simply returns the highest-scoring candidate based on tiered fitness scoring (blame $>$ blame\_ancestor $>$ bfc\_ancestor) without any LLM interaction. This isolates the TKG's candidate expansion from intelligent agent reasoning.

    \item \textit{Agent-only}: Uses an LLM agent to reason over blame candidates (with fallback) but without constructing a TKG. The agent receives blame candidate commit SHAs and their commit messages as text input and uses a simplified prompt without tool-calling capabilities, directly outputting its prediction. This isolates the agent's reasoning capability without the expanded search space provided by the TKG.

    \item \textit{\appname (w/o expansion)}: Combines TKG construction with agent-guided search but blames only the exact deleted or modified lines as seed candidates, omitting the two-line context expansion around each blamed line introduced in Section~\ref{subsec:tkg-construction}. This isolates the contribution of expanding the blame seed set beyond the lines directly modified by the BFC.

    \item \textit{Full Pipeline (\appname)}: Combines TKG construction with agent-guided search and applies the two-line context expansion around each blamed line to widen the blame seed set. The agent navigates the TKG using the four specialized tools (Section~\ref{subsec:search-agent}) for candidate enumeration, graph traversal, property queries, and causal analysis between BFC and BIC candidates.
\end{enumerate}

By comparing these configurations, we quantify each component's marginal contribution and assess whether their combination produces synergistic benefits. Since in RQ1, we observed at most a variation of 3\% for any metric across the three runs of either AgenticSZZ or LLM4SZZ, for RQ2's analysis, we decided to only perform one run of each variant.

\subsubsection{Results}\label{subsubsec:rq2-results}

% \hao{storyline:
%     Part 1: full pipeline is the best, explain why and insights
%     Part 2: Blame-only is the floor. Blame-fallback proves that handling blameless cases matters (connect to RQ0). TKG-only (deterministic) shows that pure structural expansion seems noisy if we don't have intelligence (agent/llm) to filter it.
%     Part 3: Acknowledge Agent-only beats TKG-only, but Full is the value that TKG brings in. So the Agent is smart (can compare with LLM4SZZ here), but it hits a ceiling without the TKG (our core contribution).}
% fixed, revised to follow storyline, avoid repeating numbers

\begin{table}[t]
    \centering
    \caption{Ablation study of \appname components. $^{*}$marks F1 differences where \appname is statistically significantly better than all four basic ablations ($p<0.05$, McNemar's test).}
    \label{tab:rq2-ablation}
    \footnotesize
    \begin{tabular}{l rrr rrr rrr}
        \toprule
                                  & \multicolumn{3}{c}{\textbf{DS\_LINUX}} & \multicolumn{3}{c}{\textbf{DS\_GITHUB}} & \multicolumn{3}{c}{\textbf{DS\_APACHE}}                                                                                                             \\
        \cmidrule(lr){2-4} \cmidrule(lr){5-7} \cmidrule(lr){8-10}
        \textbf{Ablations}        & Prec.                                  & Recall                                  & F1-score                                & Prec.          & Recall         & F1-score             & Prec.          & Recall         & F1-score       \\
        \midrule
        Blame-only                & 0.491                                  & 0.471                                   & 0.481                                   & 0.591          & 0.589          & 0.590                & 0.546          & 0.374          & 0.444          \\
        Blame-fallback            & 0.555                                  & 0.533                                   & 0.544                                   & 0.632          & 0.631          & 0.631                & 0.558          & 0.383          & 0.454          \\
        TKG-only                  & 0.544                                  & 0.523                                   & 0.533                                   & 0.485          & 0.483          & 0.484                & 0.400          & 0.274          & 0.325          \\
        Agent-only                & 0.571                                  & 0.549                                   & 0.560                                   & 0.599          & 0.597          & 0.598                & 0.517          & 0.354          & 0.420          \\
        \appname\ (w/o expansion) & 0.670                                  & 0.644                                   & 0.657                                   & 0.722          & 0.720          & 0.721                & \textbf{0.608} & \textbf{0.417} & \textbf{0.495} \\
        \textbf{\appname}         & \textbf{0.695}                         & \textbf{0.668}                          & \textbf{0.681}$^{*}$                    & \textbf{0.780} & \textbf{0.778} & \textbf{0.779}$^{*}$ & 0.571          & 0.391          & 0.464          \\
        \bottomrule
    \end{tabular}
\end{table}

\textbf{The full pipeline outperforms all basic ablation configurations across all datasets, while context expansion shows dataset-dependent effects.}
Table~\ref{tab:rq2-ablation} presents results across the six configurations. \appname outperforms Blame-only, Blame-fallback, TKG-only, and Agent-only on every dataset, achieving the highest F1-score on DS\_LINUX and DS\_GITHUB; the largest margin appears on DS\_GITHUB where diverse project structures benefit most from TKG-based exploration. Comparing against the \appname (w/o expansion) variant isolates the effect of the two-line context expansion introduced in Section~\ref{subsec:tkg-construction}: expansion improves F1 by 0.024 on DS\_LINUX and 0.058 on DS\_GITHUB (both statistically significant; $p<0.01$, McNemar's test), but marginally reduces it on DS\_APACHE (0.495 to 0.464; $p=0.093$, not significant), where blame complexity is highest (3.3 blame commits per case on average). McNemar's tests confirm that \appname's F1 improvements over the four basic ablations are statistically significant on all datasets ($p<0.05$) with large effect sizes (Cohen's $g$ from 0.32 to 0.38)~\cite{cohen2013statistical}.

% \hao{Discuss Blame-only vs. Blame-fallback vs. TKG. These are all deterministic/rule-based, list pros and cons and provide insights}revised
\textbf{Among configurations without LLM reasoning, Blame-fallback provides the best baseline, while TKG-only suffers from noise introduced by search space expansion.}
Blame-only serves as the simplest baseline: it returns the most recent blame commit without handling blameless cases. Blame-fallback improves over Blame-only on all datasets by providing candidates for blameless cases, where Blame-only returns no prediction. However, TKG-only performs worse than Blame-fallback despite expanding the search space, particularly on DS\_GITHUB and DS\_APACHE. This reveals that deterministic ranking cannot effectively filter the expanded candidate set of a TKG: commits that coincidentally modified the same files are scored similarly to those that actually introduced the bug, reducing precision.

% \hao{Discuss Agent-only, mention it is based on Blame-fallback (more intelligent), then compare with TKG-only, list pros and cons}revised
\textbf{Agent-only, which operates without TKG construction, is competitive with LLM4SZZ, revealing that the TKG is the component that pushes \appname past the state of the art.}
Agent-only applies LLM reasoning over the same blame-fallback candidate set, outperforming TKG-only on all datasets and reaching performance competitive with LLM4SZZ (e.g., Linux F1=0.560 vs.\ 0.590; GitHub F1=0.598 vs.\ 0.525). This confirms that the agent design is itself competitive, but its coverage is bounded by the blame-based search space. The full pipeline closes this coverage gap by expanding the candidate set through TKG traversal, enabling \appname to consistently surpass LLM4SZZ.

\begin{table}[t]
    \centering
    \caption{True positives by BIC category. Each BFC is assigned to a single category using priority (Blame $>$ Ancestor $>$ Blameless). Ancestor combines Blame Ancestor and BFC Ancestor categories. For fair comparison, LLM4SZZ uses the same LLM (DeepSeek-V3.2) as \appname (Table~\ref{tab:llm4szz-vs-agenticszz}). Results are based on the best run (highest F1-score) for each approach.}
    \label{tab:category-breakdown}
    \footnotesize
    \begin{tabular}{l rrrr}
        \toprule
        \textbf{Approach}         & \textbf{Blame} & \textbf{Ancestor} & \textbf{Blameless} & \textbf{Total} \\
        \midrule
        Blame-only                & 1,076          & 0                 & 0                  & 1,076          \\
        Blame-fallback            & 1,076          & 0                 & 114                & 1,190          \\
        TKG-only                  & 980            & 0                 & 104                & 1,084          \\
        Agent-only                & 1,080          & 1                 & 112                & 1,193          \\
        LLM4SZZ                   & 1,017          & 36                & 123                & 1,176          \\
        \appname\ (w/o expansion) & \textbf{1,209} & 19                & 180                & 1,408          \\
        \appname                  & 1,161          & \textbf{125}      & \textbf{182}       & \textbf{1,468} \\
        \bottomrule
    \end{tabular}
\end{table}

\begin{figure}[t]
    \centering
    \begin{subfigure}[b]{0.32\textwidth}
        \includegraphics[width=\textwidth]{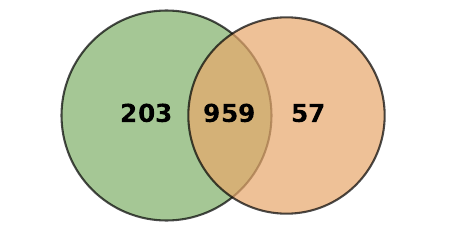}
        \caption{Blame}
    \end{subfigure}
    \hfill
    \begin{subfigure}[b]{0.32\textwidth}
        \includegraphics[width=\textwidth]{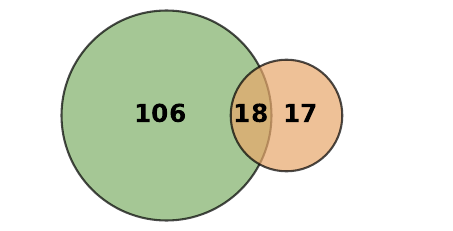}
        \caption{Ancestor}
    \end{subfigure}
    \hfill
    \begin{subfigure}[b]{0.32\textwidth}
        \includegraphics[width=\textwidth]{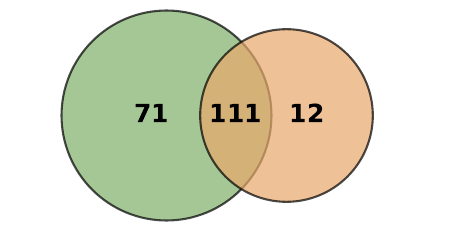}
        \caption{Blameless}
    \end{subfigure}
    \caption{True positives overlap by BIC category. Green: \appname only; Orange: LLM4SZZ only.}
    \label{fig:venn-diagrams}
\end{figure}

\textbf{BIC category analysis reveals that context expansion and TKG traversal are complementary mechanisms, with the largest gain in ancestor detection.} Table~\ref{tab:category-breakdown} and Figure~\ref{fig:venn-diagrams} break down the true positives of each approach by BIC category. \appname achieves the highest counts in Ancestor (125 vs.\ 36) and Blameless (182 vs.\ 123), and the highest total (1,468 vs.\ 1,176). The ablation results make the Ancestor mechanism explicit: the agent without TKG (Agent-only) correctly identifies only 1 Ancestor case, and \appname (w/o expansion) reaches 19, LLM4SZZ, which applies context expansion without TKG traversal, reaches 36. Combining both mechanisms yields 125 Ancestor true positives, a $3.5\times$ improvement over LLM4SZZ. Expansion widens the blame seed set while TKG traversal makes the widened space navigable; neither mechanism alone unlocks the Ancestor category.

This comes at a modest cost in the Blame category, where \appname (w/o expansion) captures more true positives (1,209) than the full pipeline (1,161), indicating that expansion trades roughly 48 Blame true positives for 106 additional Ancestor ones, a net gain of 60 overall. The Venn diagrams in Figure \ref{fig:venn-diagrams} show that \appname and LLM4SZZ are largely complementary for the Blame category (each captures cases the other misses), while the Ancestor solutions are nearly disjoint, consistent with TKG traversal exploring structural commit relationships that lie outside the linear context window LLM4SZZ expands into.

\textbf{The TKG-agent synergy reflects an exploration-exploitation trade-off~\cite{sutton2018reinforcement}.} The TKG expands the frontier of possible solutions by encoding temporal and code-level relationships, while the agent exploits semantic understanding to select among them. Neither component alone is sufficient: expansion without intelligence introduces noise, while intelligence without expansion hits a coverage ceiling. This complementary design suggests that future work could explore integrating search-based software engineering techniques~\cite{DBLP:journals/csur/HarmanMZ12} to further optimize the navigation strategy.

\begin{summarybox}{Summary of RQ2}
    \small
    \begin{itemize}[leftmargin=*]
        \item Both components (TKG construction and agent reasoning) and the context expansion strategy contribute complementary gains; the full pipeline yields the highest total true positives (1,468 vs.\ LLM4SZZ's 1,176).
        \item Context expansion and TKG traversal jointly unlock the Ancestor category: TKG traversal without expansion reaches 19 Ancestor true positives and expansion without TKG reaches 36; combined, they reach 125 (a $3.5\times$ improvement over LLM4SZZ).
    \end{itemize}
    \textbf{Implication:} The TKG-agent synergy reflects an exploration-exploitation trade-off: TKG traversal without intelligent selection introduces noise, while the agent without TKG hits a coverage ceiling. Context expansion amplifies ancestor discovery by trading 48 Blame true positives for 106 Ancestor ones, producing a net gain on DS\_LINUX and DS\_GITHUB but not on DS\_APACHE, where high blame complexity already challenges candidate selection.
\end{summarybox}

\subsection{RQ3: \rqthree}\label{subsec:rq3}

\subsubsection{Motivation}\label{subsubsec:rq3-motivation}

RQ1 evaluated \appname's effectiveness using DeepSeek-V3.2 as the sole LLM, leaving open whether these improvements generalize to other models. \appname's TKG-navigation task demands reliable multi-step tool invocation and causal code reasoning, capabilities that vary across LLMs. We focus specifically on open-weight models, which researchers can self-host to fully reproduce \appname's results without depending on proprietary API access.

\subsubsection{Approach}\label{subsubsec:rq3-approach}

We evaluate \appname's full pipeline using five open-weight LLMs covering both dense and MoE architectures:

\begin{enumerate}[leftmargin=*]
    \item \textit{DeepSeek-V3.2}~\cite{DBLP:journals/corr/abs-2512-02556} (MoE, 685B total / 37B active): the primary LLM described in Section~\ref{subsec:implementation}.
    \item \textit{DeepSeek-V4-Pro}~\cite{deepseekai2026deepseekv4} (MoE, 1.6T total / 49B active): a recently released frontier reasoning model.
    \item \textit{Gemma-4-31B-IT}~\cite{google_gemma4_2025} (dense, 31B): a multimodal instruction-tuned model.
    \item \textit{Devstral-Small-2}~\cite{mistral_devstral_2025} (dense, 24B): a code-specialized model.
    \item \textit{Qwen3-Coder-Next}~\cite{qwen_qwen3_coder_next_tech_report} (MoE, 80B total / 3B active): a code-specialized model.
\end{enumerate}

All five models are open-weight; Gemma-4-31B-IT, Devstral-Small-2, and Qwen3-Coder-Next are served via a local OpenAI-compatible inference endpoint, while DeepSeek-V3.2 and DeepSeek-V4-Pro are accessed via the official API. DeepSeek-V4-Pro is used in non-thinking mode for fair comparison. We conduct one run per LLM per dataset; for DeepSeek-V3.2, we reuse the RQ2 ablation run, as RQ1 established run-to-run variance within 3\% across all metrics. Additionally, since DeepSeek-V4-Pro~\cite{deepseekai2026deepseekv4} was released close to submission, we present a controlled comparison against LLM4SZZ using DeepSeek-V4-Pro to examine whether \appname's TKG architecture amplifies the benefits of a stronger LLM.

\subsubsection{Results}\label{subsubsec:rq3-results}

\textbf{\appname's effectiveness varies substantially with LLM choice, with a largely consistent performance ordering across datasets.} Table~\ref{tab:rq3-sensitivity} presents results across all five LLMs and three datasets. DeepSeek-V4-Pro achieves the highest F1-score on all datasets, followed by DeepSeek-V3.2 and Gemma-4-31B-IT. Devstral-Small-2 and Qwen3-Coder-Next occupy the bottom tier with similar scores on most datasets, though their relative ordering varies. The consistent top-three ranking across datasets indicates that this variation reflects general LLM capability for TKG navigation rather than dataset-specific characteristics.

\textbf{Only the two largest models (37B+ active parameters) consistently outperform the Blame-fallback baseline, revealing a minimum capability threshold for effective TKG navigation.} Comparing against the Blame-fallback baseline included in Table~\ref{tab:rq3-sensitivity}, both DeepSeek-V4-Pro and DeepSeek-V3.2 exceed it on all three datasets. Gemma-4-31B-IT surpasses Blame-fallback only on DS\_LINUX (0.582 vs. 0.544) but falls below it on DS\_GITHUB (0.534 vs. 0.631) and DS\_APACHE (0.410 vs. 0.454). Devstral-Small-2 and Qwen3-Coder-Next fall below Blame-fallback on all datasets, with both achieving similar F1-scores on DS\_LINUX and DS\_GITHUB despite an 8$\times$ difference in active parameters. This reveals that smaller models make fewer tool calls per case (Table~\ref{tab:rq3-tool-use}), skipping the multi-step verification that effective TKG navigation requires and making the simple Blame-fallback heuristic a safer choice for those models.

\textbf{Tool-use analysis reveals that the capability gap stems from insufficient candidate verification.} Table~\ref{tab:rq3-tool-use} shows per-case tool invocation rates (mean) across LLMs. Most models call \texttt{list\_candidates} approximately once to retrieve the candidate overview, though Qwen3-Coder-Next averages 1.58 calls, suggesting repeated exploration rather than progression. Models diverge sharply on subsequent verification. DeepSeek-V4-Pro calls \texttt{query\_node} 1.58 times per case and reads 2.78 diffs, while DeepSeek-V3.2 calls it 0.90 times and reads 2.14 diffs. In contrast, smaller models nearly skip metadata verification: Gemma-4-31B-IT calls \texttt{query\_node} only 0.04 times and reads just 1.25 diffs per case. \texttt{traverse\_graph} is used sparingly by all models (0.00 to 0.22), indicating that graph traversal plays a supporting rather than primary role. This pattern explains why smaller models fall below the Blame-fallback baseline: they accept the initial candidate ranking from \texttt{list\_candidates} without the verification steps needed to correct ranking errors, effectively reducing the agent to a less reliable heuristic.

% RQ3 LLM sensitivity table
\begin{table*}[t]
    \centering
    \caption{LLM sensitivity analysis: \appname full pipeline performance across five LLMs (one run each). DeepSeek-V3.2 reuses the RQ2 ablation run.}
    \label{tab:rq3-sensitivity}
    \footnotesize
    \begin{tabular}{l rrr rrr rrr}
        \toprule
        \multirow{2}{*}{\textbf{LLM}} & \multicolumn{3}{c}{\textbf{DS\_LINUX}} & \multicolumn{3}{c}{\textbf{DS\_GITHUB}} & \multicolumn{3}{c}{\textbf{DS\_APACHE}}                                                 \\
        \cmidrule(lr){2-4} \cmidrule(lr){5-7} \cmidrule(lr){8-10}
                                      & Prec.                                  & Rec.                                    & F1                                      & Prec. & Rec.  & F1    & Prec. & Rec.  & F1    \\
        \midrule
        Blame-fallback                & 0.555                                  & 0.533                                   & 0.544                                   & 0.632 & 0.631 & 0.631 & 0.558 & 0.383 & 0.454 \\
        \midrule
        DeepSeek-V3.2                 & 0.695                                  & 0.668                                   & 0.681                                   & 0.780 & 0.778 & 0.779 & 0.571 & 0.391 & 0.464 \\
        \textbf{DeepSeek-V4-Pro}      & 0.795                                  & 0.764                                   & 0.779                                   & 0.811 & 0.808 & 0.809 & 0.579 & 0.397 & 0.471 \\
        Gemma-4-31B-IT                & 0.594                                  & 0.571                                   & 0.582                                   & 0.535 & 0.533 & 0.534 & 0.504 & 0.346 & 0.410 \\
        Devstral-Small-2              & 0.508                                  & 0.488                                   & 0.498                                   & 0.446 & 0.444 & 0.445 & 0.442 & 0.303 & 0.359 \\
        Qwen3-Coder-Next              & 0.517                                  & 0.497                                   & 0.507                                   & 0.446 & 0.444 & 0.445 & 0.362 & 0.249 & 0.295 \\
        \bottomrule
    \end{tabular}
\end{table*}

% RQ3 tool-use table
\begin{table}[t]
    \centering
    \caption{Agent tool-use patterns across LLMs (average calls per case). \textit{Steps}: median total tool calls per case.}
    \label{tab:rq3-tool-use}
    \footnotesize
    \begin{tabular}{l r rrrr}
        \toprule
        \textbf{LLM}     & \textbf{Steps} & \texttt{list\_c.} & \texttt{query\_n.} & \texttt{read\_n.} & \texttt{trav\_g.} \\
        \midrule
        DeepSeek-V3.2    & 5              & 1.00              & 0.90               & 2.14              & 0.20              \\
        DeepSeek-V4-Pro  & 4              & 1.01              & 1.58               & 2.78              & 0.22              \\
        Gemma-4-31B-IT   & 2              & 1.10              & 0.04               & 1.25              & 0.00              \\
        Devstral-Small-2 & 3              & 1.00              & 0.18               & 1.90              & 0.13              \\
        Qwen3-Coder-Next & 3              & 1.58              & 0.29               & 1.99              & 0.19              \\
        \bottomrule
    \end{tabular}
\end{table}

\textbf{\appname's architecture amplifies the benefits of stronger LLMs, while LLM4SZZ's performance plateaus.} Table~\ref{tab:llm4szz-vs-agenticszz-v4pro} presents a controlled comparison where both approaches use DeepSeek-V4-Pro (single run). Compared to the V3.2 results (Table~\ref{tab:llm4szz-vs-agenticszz}), \appname's overall F1-score improves from 0.669 to 0.751 (+8.2 percentage points), while LLM4SZZ improves only marginally from 0.563 to 0.569 (+0.6 percentage points). The relative improvement of \appname over LLM4SZZ consequently widens from 18.8\% to 32.0\%. Per-dataset, the largest gains appear on DS\_LINUX (F1: 0.682 to 0.779) and DS\_GITHUB (0.781 to 0.809); DS\_APACHE remains comparable (0.468 vs.\ 0.471), consistent with the blame complexity challenges observed throughout this evaluation.

This asymmetric scaling reflects a fundamental architectural difference: \appname's TKG provides a structured reasoning space that a stronger LLM can navigate more effectively, while LLM4SZZ's performance is bounded by its blame-based candidate set regardless of LLM capability. Even with DeepSeek-V3.2, \appname already outperforms LLM4SZZ; upgrading to DeepSeek-V4-Pro only increases that gap. This finding suggests that \appname is well-positioned to benefit from future LLM improvements, whereas approaches confined to blame-based search spaces face an inherent performance ceiling.

\begin{table*}[t]
    \centering
    \caption{Controlled comparison with DeepSeek-V4-Pro: both approaches use the same LLM (single run). Compared to the V3.2 baseline (Table~\ref{tab:llm4szz-vs-agenticszz}), \appname's advantage widens from 18.8\% to 32.0\% in overall F1.}
    \label{tab:llm4szz-vs-agenticszz-v4pro}
    \footnotesize
    \begin{tabular}{l rrr rrr rrr rrr}
        \toprule
        \multirow{2}{*}{\textbf{Approach}} & \multicolumn{3}{c}{\textbf{DS\_LINUX}} & \multicolumn{3}{c}{\textbf{DS\_GITHUB}} & \multicolumn{3}{c}{\textbf{DS\_APACHE}} & \multicolumn{3}{c}{\textbf{Overall}}                                                                                                                                         \\
        \cmidrule(lr){2-4} \cmidrule(lr){5-7} \cmidrule(lr){8-10} \cmidrule(lr){11-13}
                                           & Prec.                                  & Rec.                                    & F1                                      & Prec.                                & Rec.           & F1             & Prec.          & Rec.           & F1             & Prec.          & Rec.           & F1             \\
        \midrule
        LLM4SZZ                            & 0.601                                  & 0.597                                   & 0.599                                   & 0.594                                & 0.461          & 0.519          & 0.536          & \textbf{0.421} & 0.472          & 0.592          & 0.548          & 0.569          \\
        \appname                           & \textbf{0.795}                         & \textbf{0.764}                          & \textbf{0.779}                          & \textbf{0.811}                       & \textbf{0.808} & \textbf{0.809} & \textbf{0.579} & 0.397          & \textbf{0.471} & \textbf{0.773} & \textbf{0.730} & \textbf{0.751} \\
        \bottomrule
    \end{tabular}
\end{table*}

\begin{summarybox}{Summary of RQ3}
    \small
    \begin{itemize}[leftmargin=*]
        \item \appname's effectiveness varies substantially with LLM choice; DeepSeek-V4-Pro achieves the highest F1-score on all datasets, with the top-three ranking (V4-Pro, V3.2, Gemma) consistent across all datasets.
        \item Only the two largest models consistently outperform the Blame-fallback baseline; smaller models make fewer tool calls per case, skipping multi-step candidate verification.
        \item Tool-use analysis shows the gap stems from insufficient verification: smaller models largely skip \texttt{query\_node} metadata checks (0.04 to 0.29 calls vs.\ 0.90 to 1.58 for the two largest models), accepting initial rankings without the checks needed to correct errors.
        \item A controlled comparison using DeepSeek-V4-Pro shows that \appname's TKG architecture amplifies stronger LLMs (F1 gap over LLM4SZZ widens from 18.8\% to 32.0\%), while LLM4SZZ's blame-bounded search space limits its gains.
    \end{itemize}
    \textbf{Implication:} \appname requires a sufficiently capable LLM for reliable multi-step TKG navigation; practitioners should validate that their chosen LLM exceeds the Blame-fallback baseline before deployment.
\end{summarybox}

\section{Discussion and Future Work}\label{sec:discussion}

% \subsection{Controlled Comparison with LLM4SZZ}\label{subsec:comparison}
% \hao{move to RQ1?}DONE

% \subsection{The TKG-Agent Synergy}\label{subsec:synergy}
% \hao{move (part or full) to RQ2?}DONE - moved key insight to RQ2 Results
% Depending on pages, we can add more cost analysis discussion here later, if not over 18 pages
\subsection{Project-Level Patterns Affecting BIC Identification}\label{subsec:project-patterns}

% Our preliminary study (RQ0) characterized the distribution of blame complexity across datasets, which we hypothesized to be an important influence for BIC identification difficulty.

\textbf{Blame complexity varies across datasets.} As shown in Figure~\ref{fig:complexity-heatmap}, projects differ substantially in their blame complexity. DS\_APACHE averages 3.3 blame commits per case, compared to 1.7 for DS\_LINUX and 2.2 for DS\_GITHUB. This correlates with our observation that \appname's improvement on DS\_APACHE is smaller and not statistically significant ($p=1.000$), while improvements on DS\_LINUX and DS\_GITHUB are significant. While our evaluation did not directly measure the relationship between blame count and accuracy, this pattern suggests that blame complexity could serve as an indicator of case difficulty, warranting future investigation.

\textbf{Simple cases may benefit from fast-path heuristics.} Projects with predominantly single-blame cases (e.g., DS\_LINUX where 47\% have 1 to 2 deleted lines with 1 blame commit) may benefit from fast-path optimizations that resolve cases with minimal LLM interaction. Future work should systematically evaluate accuracy stratified by blame complexity to confirm this hypothesis.

\subsection{Practical Considerations}\label{subsec:practical}

\textbf{Cost-effectiveness.} \appname requires LLM API calls for agent reasoning, incurring computational costs not present in traditional SZZ variants. However, as shown in Table~\ref{tab:cost-average}, the average cost per case remains low (\$0.004 to \$0.005) due to DeepSeek-V3.2's competitive pricing. For contexts where BIC identification directly impacts downstream tasks such as defect prediction model training or automated program repair, this cost may be justified by the improved accuracy.

\textbf{Integration with existing workflows.} The TKG construction phase operates entirely on local repository data and can be cached across multiple queries to the same repository. This makes \appname suitable for integration into CI/CD pipelines where the same repository is analyzed repeatedly, amortizing the graph construction cost.

\subsection{Concurrent Agentic SZZ Approaches and Future Work}\label{subsec:concurrent}

\textbf{\appname is the first agentic approach to BIC identification, with two independent concurrent works appearing approximately two months after our initial arXiv submission (February 2026).} Two concurrent works, SZZ-Agent~\cite{risse2026and} (March 2026) and AgentSZZ~\cite{lyu2026agentszz} (April 2026), provide convergent evidence that agentic paradigms represent a genuine advance for BIC identification. The key conceptual difference is structure: while our \appname constructs a Temporal Knowledge Graph encoding commits with temporal and structural relationships, SZZ-Agent uses binary search over raw file history and AgentSZZ employs open-ended repository exploration via git tools. Both concurrent works rely on closed-source LLMs from Anthropic and OpenAI (e.g., the Claude and GPT series), while \appname uses open-weight DeepSeek-V3.2 supporting reproducible evaluation. SZZ-Agent additionally evaluates on sampled subsets rather than complete datasets. Beyond BIC identification, MAS-SZZ~\cite{cao2026masszzmultiagenticszzalgorithm} extends the agentic paradigm to vulnerability-inducing commit identification, using multi-agent collaboration with CVE descriptions as additional input. A comprehensive comparison across agentic approaches on identical datasets with standardized open-weight LLMs is an important direction for future work.

Several directions for future work emerge from our empirical findings. First, the persistent challenges on high-complexity cases (e.g., DS\_APACHE with 3.3 blame commits per case) suggest that incorporating code semantic similarity or developer expertise patterns could help disambiguate among competing candidates. Second, while we focused on single-BIC prediction, extending \appname to identify multiple co-dependent BICs would better handle cases where bugs emerge from the interaction of several commits. Third, exploring alternative graph schemas that capture semantic code relationships (e.g., data flow, call graphs) beyond file and function co-modification could enable more precise causal reasoning. Fourth, our LLM sensitivity analysis shows that smaller models fall below the Blame-fallback baseline, raising an open question of whether fine-tuning, distillation, or model-specific prompt engineering could lower the capability threshold and make TKG-guided search accessible with smaller, locally deployable models. Our evaluation uses a single prompt across all LLMs for fair comparison.

\section{Threats to Validity}\label{sec:threats}

% \subsection{Internal Validity}\label{sec:InternalValidity}

\textbf{Internal Validity.}
% \textbf{LLM Selection.}
\appname uses DeepSeek-V3.2 while LLM4SZZ employs GPT-4o-mini. We selected DeepSeek-V3.2 because it demonstrated more reliable tool invocation in our preliminary experiments, which is essential for TKG navigation. The controlled comparison in RQ1 (Table~\ref{tab:llm4szz-vs-agenticszz}) confirms that our improvements over LLM4SZZ stem from the TKG-guided paradigm rather than a more modern LLM version. RQ3 further addresses LLM sensitivity by evaluating \appname across five open-weight models, finding that only the two largest models consistently surpass the Blame-fallback baseline across all datasets, revealing a minimum capability threshold for effective TKG navigation. Additionally, DeepSeek-V4-Pro was evaluated in non-thinking mode for fair comparison with the other models; enabling its reasoning mode may further improve results, but due to cost constraints we leave this for future work.

\begin{figure}[t]
    \centering
    \begin{subfigure}[b]{0.32\textwidth}
        \includegraphics[width=\textwidth]{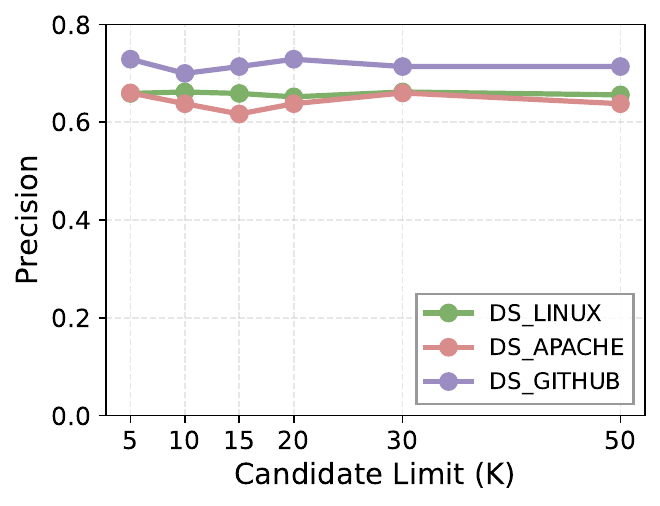}
    \end{subfigure}
    \hfill
    \begin{subfigure}[b]{0.32\textwidth}
        \includegraphics[width=\textwidth]{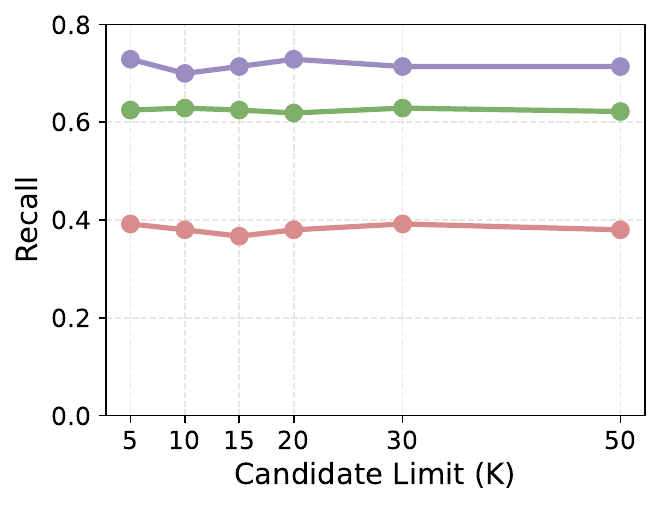}
    \end{subfigure}
    \hfill
    \begin{subfigure}[b]{0.32\textwidth}
        \includegraphics[width=\textwidth]{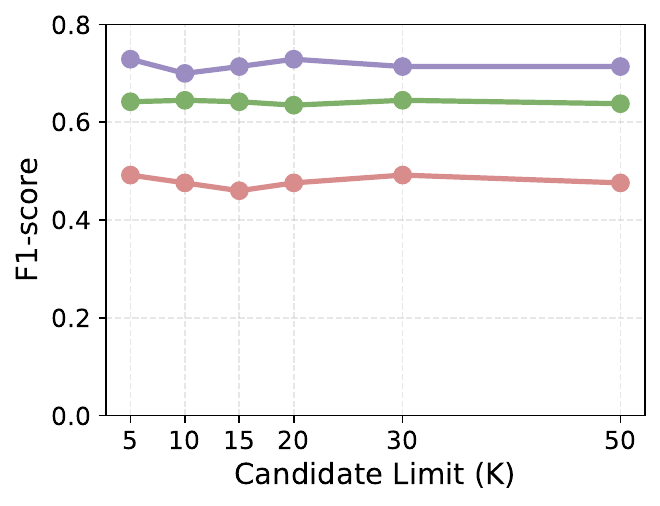}
    \end{subfigure}
    \caption{Sensitivity of \appname performance to candidate limit $K$ across precision, recall, and F1-score.}
    \label{fig:sensitivity-k}
\end{figure}

% \textbf{Sensitivity Analysis.}
\appname uses a candidate limit parameter $K$ that controls the maximum number of BIC candidates retrieved from the TKG for agent evaluation. Figure~\ref{fig:sensitivity-k} examines the extent to which our results are sensitive to this parameter choice; due to the computational cost of running six $K$ values across all 2,102 cases, we conduct this sensitivity experiment on a 20\% stratified evaluation sample ($N$=416; DS\_LINUX: 299, DS\_GITHUB: 70, DS\_APACHE: 47). Across $K$ values from 5 to 50, performance remains remarkably stable: DS\_LINUX F1-score varies within 1.0\% (0.635 to 0.645), DS\_GITHUB within 2.9\% (0.700 to 0.729), and DS\_APACHE within 3.2\% (0.460 to 0.492). This stability indicates that the agent's reasoning effectively filters candidates regardless of initial pool size, and our reported results do not need to rely on a carefully tuned $K$ value. Even aggressive filtering ($K$=5) achieves near-optimal performance, suggesting that TKG ranking places true BICs near the top of candidate lists.

% \hao{three sub figures, y-axis starts from 0}fixed

% \subsection{External Validity}\label{sec:ExternalValidity}
\smallskip\noindent\textbf{External Validity.}
Our evaluation uses three datasets (DS\_LINUX, DS\_APACHE, DS\_GITHUB) comprising C and Java projects. While these represent diverse development practices, from the highly-structured Linux kernel to varied open-source DS\_APACHE and GitHub projects, our findings may not generalize to other languages (e.g., Python, JavaScript) or proprietary codebases with different commit conventions. The datasets contain developer-annotated BIC labels, which provide higher ground truth quality but may reflect annotation biases. Future work should evaluate \appname on additional languages and project types.

\section{Conclusion}\label{sec:conclusion}

We presented \appname, a novel approach that reframes Bug-Inducing Commit (BIC) identification from a ranking problem over blame commits into a graph search problem over a Temporal Knowledge Graph. Our preliminary study (RQ0) on 2,102 validated cases revealed that 28\% of BICs require traversal beyond direct blame commits and 14\% of cases are blameless, motivating the need for expanded search capabilities. \appname addresses these challenges through a two-phase architecture: TKG construction that encodes temporal and structural relationships among commits, and an LLM agent that navigates this graph using specialized tools for candidate enumeration, structural traversal, and causal analysis.

Our evaluation demonstrates that \appname outperforms all baselines on three of four dataset splits, achieving F1-scores of 0.47 to 0.79 across datasets and improving over state-of-the-art by 16.0\% to 34.2\% in F1. The ablation study confirms that both components and the context expansion strategy contribute: TKG and agent form a necessary exploration-exploitation pair, and context expansion jointly unlocks the Ancestor category, yielding a net gain of 60 true positives across datasets. Controlled comparisons with LLM4SZZ using identical LLMs, both DeepSeek-V3.2 and the newer V4-Pro, validate that our improvements stem from the TKG-guided search paradigm rather than LLM selection; upgrading the backbone LLM widens \appname's advantage (F1 gap from 18.8\% to 32.0\%) because the TKG architecture amplifies stronger reasoning, whereas LLM4SZZ's blame-bounded search space limits its gains. Our LLM sensitivity analysis further reveals that effective TKG navigation requires a sufficiently capable model; smaller models make fewer tool calls per case, skipping multi-step candidate verification and falling below the simple Blame-fallback heuristic.

% \textbf{Broader Impact.} By reframing BIC identification from a ranking problem into a graph search problem, we open a new research direction that invites techniques from search-based software engineering and graph neural networks. More accurate BIC identification directly benefits downstream tasks: defect prediction models trained on cleaner BIC labels should achieve higher accuracy, and history-aware automated program repair systems~\cite{DBLP:journals/corr/abs-2501-09135, DBLP:journals/corr/abs-2511-01047} can leverage correct BICs to provide precise historical context, enabling more targeted and effective patches.

% \section{Acknowledgments}

% Identification of funding sources and other support, and thanks to
% individuals and groups that assisted in the research and the
% preparation of the work should be included in an acknowledgment
% section, which is placed just before the reference section in your
% document.

% This section has a special environment:
% \begin{verbatim}
%   \begin{acks}
%   ...
%   \end{acks}
% \end{verbatim}

% looks this will be hidden when we use the required \documentclass[acmsmall,screen,review,anonymous]{acmart}
% \begin{acks}
%     This research is supported ......
% \end{acks}

% need to comment out for arXiv Submitted Version
\section*{Data Availability}
The replication package, including source code, datasets, and evaluation scripts, is available at 
\url{https://github.com/SAILResearch/AgenticSZZ}.
% \url{https://anonymous.4open.science/r/AgenticSZZ-Replication}.

\bibliographystyle{ACM-Reference-Format}
\bibliography{sample-base}

%%
%% System prompt appendix moved to replication package to save space.
%% The complete prompt is available in the replication package at:
%% agentic_szz/agent/bic_search_agent.py (SYSTEM_PROMPT variable)
\appendix

\section{Agent System Prompt}
\label{appendix:prompt}

The following shows the complete system prompt used by the \appname agent. The prompt includes practical heuristic guidance (e.g., fast-path rules for simple cases) that instantiate the fitness-based ranking strategy described in Section~\ref{subsec:search-agent}. These heuristics help the agent make efficient decisions while the underlying approach remains the type-based fitness scoring.

\begin{tcolorbox}[
        colback=gray!5!white,
        colframe=gray!75!black,
        title=System Prompt,
        fonttitle=\bfseries\small,
        breakable,
        enhanced,
        left=3mm,
        right=3mm,
        top=2mm,
        bottom=2mm
    ]
    \small
    You are a Bug-Inducing Commit (BIC) identification expert.

    \medskip
    \textbf{Task:} Given a Bug-Fixing Commit (BFC), identify which commit first INTRODUCED the bug.

    \medskip
    \textbf{Commit Types (Priority Order):}
    \begin{enumerate}
        \item \textbf{blame}: Code deleted/modified by BFC was last touched by this commit. HIGHEST priority.
        \item \textbf{blame\_ancestor}: Commit BEFORE blame in file history. Lower priority.
        \item \textbf{bfc\_ancestor}: Commit AFTER blame but BEFORE BFC. Lowest priority.
    \end{enumerate}

    \medskip
    \textbf{CRITICAL: Fast-Path Rules (MUST FOLLOW)}

    \textit{IF ``SINGLE BLAME COMMIT'' appears:}
    \begin{itemize}
        \item STOP IMMEDIATELY. Return that commit.
        \item Do NOT call any other tools. Do NOT investigate alternatives.
    \end{itemize}

    \textit{IF ``DOMINANT COMMIT'' or ``TOP CANDIDATE'' with $>$70\% appears:}
    \begin{itemize}
        \item This is the LIKELY BIC based on blame line analysis.
        \item Return it UNLESS you have specific evidence another candidate is better.
        \item Check other candidates only if uncertain.
    \end{itemize}

    \textit{IF ``TOP CANDIDATE'' is marked:}
    \begin{itemize}
        \item This candidate has the highest fitness score based on code overlap.
        \item Return it unless you find strong evidence it's wrong.
    \end{itemize}

    \medskip
    \textbf{Multi-Blame Cases Only:}
    If no single/dominant commit and multiple [blame] commits exist:
    \begin{enumerate}
        \item Focus on candidates marked with high fitness scores.
        \item VERIFY the top candidate using \texttt{read\_node\_content}:
              \begin{itemize}
                  \item ``Did this commit ADD the code that the BFC REMOVES?''
                  \item ``Does this commit introduce the LOGIC showing the bug?''
              \end{itemize}
        \item If the top blame candidate is just a REFACTORing or IMPORT change: REJECT it (even if fitness is high). Check [ancestor] candidates---they may contain the actual bug introduction.
        \item IMPORTANT: [ancestor] commits have lower fitness but may be the true BIC. If blame commits look like cleanup/refactoring, explore ancestors.
    \end{enumerate}

    \medskip
    \textbf{MANDATORY: Minimum Investigation}
    \begin{itemize}
        \item You MUST call at least ONE investigation tool before \texttt{final\_result}.
        \item Exception: SINGLE BLAME fast-path with explicit ``STOP!'' instruction.
        \item Use \texttt{query\_node} or \texttt{read\_node\_content} to verify your choice.
    \end{itemize}

    \medskip
    \textbf{Available Tools (TKG Graph Operations):}
    \begin{enumerate}
        \item \texttt{list\_candidates} -- Call FIRST. Enumerates nodes with fitness scores.
        \item \texttt{read\_node\_content} -- CRITICAL for verification. Reads diff stored in node.
        \item \texttt{traverse\_graph} -- Finds connected commits via File/Function edges.
        \item \texttt{query\_node} -- Access commit node properties and related facts.
    \end{enumerate}

    \medskip
    \textbf{Cost Efficiency (Diff Tool Limit):}
    \begin{itemize}
        \item LIMIT \texttt{read\_node\_content} calls to 3 maximum per case (saves cost without hurting accuracy).
        \item Prioritize: Check TOP candidate first, then only check others if genuinely uncertain.
        \item If top 2 candidates both look plausible after diff check, pick the higher-fitness one.
        \item Do NOT exhaustively diff all candidates---diminishing returns after 3 calls.
    \end{itemize}

    \medskip
    \textbf{Output:} Return the FULL 40-character commit SHA. Explain briefly why.
\end{tcolorbox}

\end{document}